\documentclass[prb,aps,superscriptaddress,reprint]{revtex4-1}
\usepackage{epsfig}
\usepackage{epsf}
\usepackage{graphicx}
\usepackage{dcolumn}
\usepackage{subfigure}
\usepackage{bm}
\usepackage{tabularx}
\usepackage[dvips]{color}
\usepackage[latin1]{inputenc}

\def\be{\begin{equation}}
\def\ee{\end{equation}}
\def\ba{\begin{array}}
\def\ea{\end{array}}
\def\bea{\begin{eqnarray}}
\def\eea{\end{eqnarray}}

\def\Map64{/home/damien/Mesoplasticity/N64}

\def\Plots3{Plots4}

\begin{document}

\title{Mechanical noise dependent Aging and Shear Banding behavior \\ of
  a mesoscopic model of amorphous plasticity}

\author{Damien Vandembroucq}
\affiliation{
Laboratoire PMMH, CNRS/ESPCI/Univ. Paris 6 UPMC/Univ. Paris 7 Diderot\\
10 rue Vauquelin, 75231 Paris cedex 05, France}
\author{St\'ephane Roux}
\affiliation{
LMT-Cachan, ENS de Cachan/CNRS-UMR 8535/Universit\'{e} Paris 6/PRES
    UniverSud Paris\\
61 Avenue du Pr\'esident Wilson, 94235 Cachan cedex, France}

\begin{abstract}
We discuss aging and localization in a simple ``Eshelby'' mesoscopic
model of amorphous plasticity. Plastic deformation is assumed to occur
through a series of local reorganizations. Using a discretization of
the mechanical fields on a discrete lattice, local reorganizations are
modeled as local slip events. Local yield stresses are randomly
distributed in space and invariant in time. Each plastic slip event
induces a long-ranged elastic stress redistribution.  Mimicking the
effect of aging, we focus on the behavior of the model when the
initial state is characterized by a distribution of high local yield
stress values. A dramatic effect on the localization behavior is
obtained: the system first spontaneously self-traps to form a shear
band which then only slowly broadens.  The higher the ``age'' parameter
the more localized the plastic strain field. Two-time correlation
computed on the stress field show a divergent correlation time with
the age parameter. The amplitude of a local slip event (the prefactor
of the Eshelby singularity) as compared to the yield stress
distribution width acts here as a mechanical effective
temperature-like parameter: the lower the slip increment, the higher
the localization and the decorrelation time.
\end{abstract}

\maketitle

\section{Introduction}

While metals are characterized by a low elastic limit and large
  deformation before failure, their amorphous counterparts, Bulk
  Metallic Glasses (BMG) are known for the exact opposite, high
  mechanical strength and low ductility. The propensity of plastic
  deformation in BMG to localize and form shear bands is the main
  mechanism leading to mechanical failure. Understanding and thus
  controlling shear band formation is the main challenge that has so
  far limited the use of glasses as structural
  materials\cite{Schuh-ActaMat07,RTV-MSMSE11} 

Although plastic deformation may be responsible for a significant
  heat production, recent experimental studies have shown that the
  origin of nucleation and propagation of shear bands could not be
  attributed to an adiabatic shear banding instability via local
  temperature rise\cite{Lewandowski-NatMat06}. In
  absence of such a thermal softening mechanism, Falk, Shi and
  collaborators have explored by Molecular Dynamics simulations the
  hypothesis of a structural softening
  mechanism\cite{Falk-PRL05,*Falk-PRB06,*Falk-PRL07}. They showed in particular
  that shear-banding was facilitated by a high degree of relaxation of
  the glassy structure.

They proposed an interpretation of this phenomenon in the
  framework of the Shear Transformation Zones (STZ)
  theory\cite{FalkLanger-PRE98} describing plastic deformation of
  metallic glasses as resulting of local inelastic
  transformations\cite{Argon-ActaMet79}. Let us recall that according
  to STZ theory plastic deformation is assumed to result from a series
  of local reorganizations occurring within a population of ``small''
  atomic/molecular clusters (zones) through
  micro-instabilities. Plastic deformation directly results from the
  balance between flips in the positive and negative directions of
  these Shear Transformation Zones at rate that depends on an
  intensive parameter (e.g. free volume or effective structural temperature).
  Shi and Falk could associate associate the shear-band with a
  structural signature characterized by an effective temperature,
  reflecting the higher potential energy in the band than in the still
  surrounding. In a similar spirit, Manning, Langer and
  collaborators\cite{Manning-PRE07,*Manning-PRE09} proposed an enriched
  version of the STZ theory able to capture strain localization. The
  introduction of a relaxation-diffusion equation of the effective structural
  temperature was in particular shown to induce shear-banding in aged
  structure (low effective structural temperature) and/or high shear-rate
  conditions.

Independently, starting from the trap model developed by
Bouchaud\cite{Bouchaud-JPI92} for the glass transition, Sollich, Cates and
Lequeux\cite{Sollich-PRL97} developed a Soft Glassy Rheology (SGR) model to
capture the rheology of complex fluids.  In the trap model a landscape of traps
of depth $E$ drawn from an exponential distribution $\exp(-E/E_0)$ is assumed. A
break-down of ergodicity naturally emerges at $T_0=E_0/k$. From this simplified
view of the glass transition, Sollich {\it et al} introduce the mechanical
stress as a bias to the energy landscape. It is important to note that the
temperature in their model is not associated to a real thermal bath but is
assumed to emerge from some mechanical noise {\it a priori} related to elastic
interactions induced by local reorganizations.

While STZ and SGR models capture part of the rich phenomenology of
amorphous visco-plasticity, their mean-field character does not allow
them to account for localization unless an additional ingredient
  is introduced. The latter can be the relaxation/diffusion of a state
  variable as discussed above and/or the inclusion of anisotropic
  elastic effect of local plastic events (Eshelby
  inclusion\cite{Eshelby57}) in the modeling.

Building on the latter
grounds several authors have developed ``Eshelby'' mesoscopic models
to study plasticity of amorphous
materials\cite{BulatovArgon94a,*BulatovArgon94b,*BulatovArgon94c,BVR-PRL02,Picard-PRE02,Picard-PRE05,Jagla-PRE07,Schuh-ActaMat09,*Homer-PRB10,TPRV-meso10,*TPVR-PRE11}.
Except in the case of Ref. \cite{Jagla-PRE07} where a state variable
is implemented or of Ref. \cite{Picard-PRE02} where the presence of
walls traps plastic deformation, in such models, localization appears
to be only transient and complex spatio-temporal correlations very
similar to those observed in atomistic simulations emerge from the
competition between diffusion and
localization\cite{Maloney-JPCM08,*Maloney-PRL09,TPRV-meso10}.

Recently Fielding and collaborators\cite{Fielding-SM09,Fielding-PRL11}
investigated an age-dependent transient shear banding behavior in
different models where the shear banding was not triggered by an
elastic or viscous softening constitutive law, but rather through an
aging/rejuvenation behavior where the diffusive character of an
internal variable would dictate the widening and progressive vanishing
of an initial shear band.  The introduction of such a mechanism in a
variant of the SGR model results in a very slow (``glassy'') spreading
of such shear bands.

This age dependence of shear banding and its fast or glassy relaxation
motivates us to reassess the question of the  connection to be made between the
glass theory inspired SGR model and the STZ model built from the identification
of the microscopic mechanism of plasticity in amorphous materials. In
particular, it has remained so far difficult to give a microscopic justification
to the effective mechanical temperature defined in the SGR
model\cite{Lemaitre-preprint06,Falk-PRL07}.

In the following we present results about aging and localization
obtained with the original mesoscopic model of plasticity presented in
details in Ref. \cite{TPRV-meso10,TPVR-PRE11}. We discuss in
particular the effect of two parameters of the model which will appear
to respectively mimic the age of the system before shearing and a
mechanical effective temperature.

\begin{figure*}[tb]
\begin{minipage}[h]{1.0\textwidth}
\includegraphics[width=0.195\textwidth]{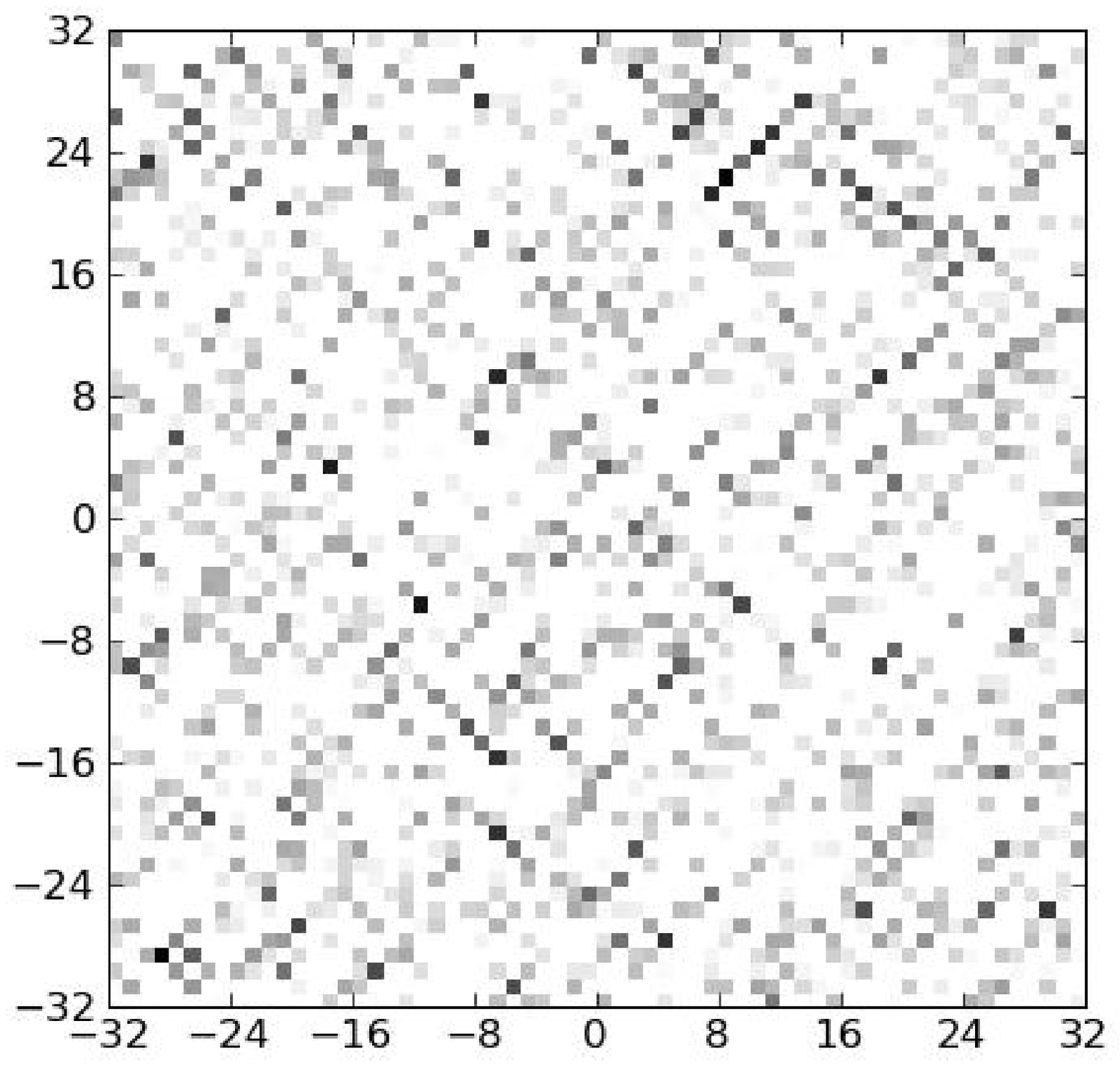}
\includegraphics[width=0.195\textwidth]{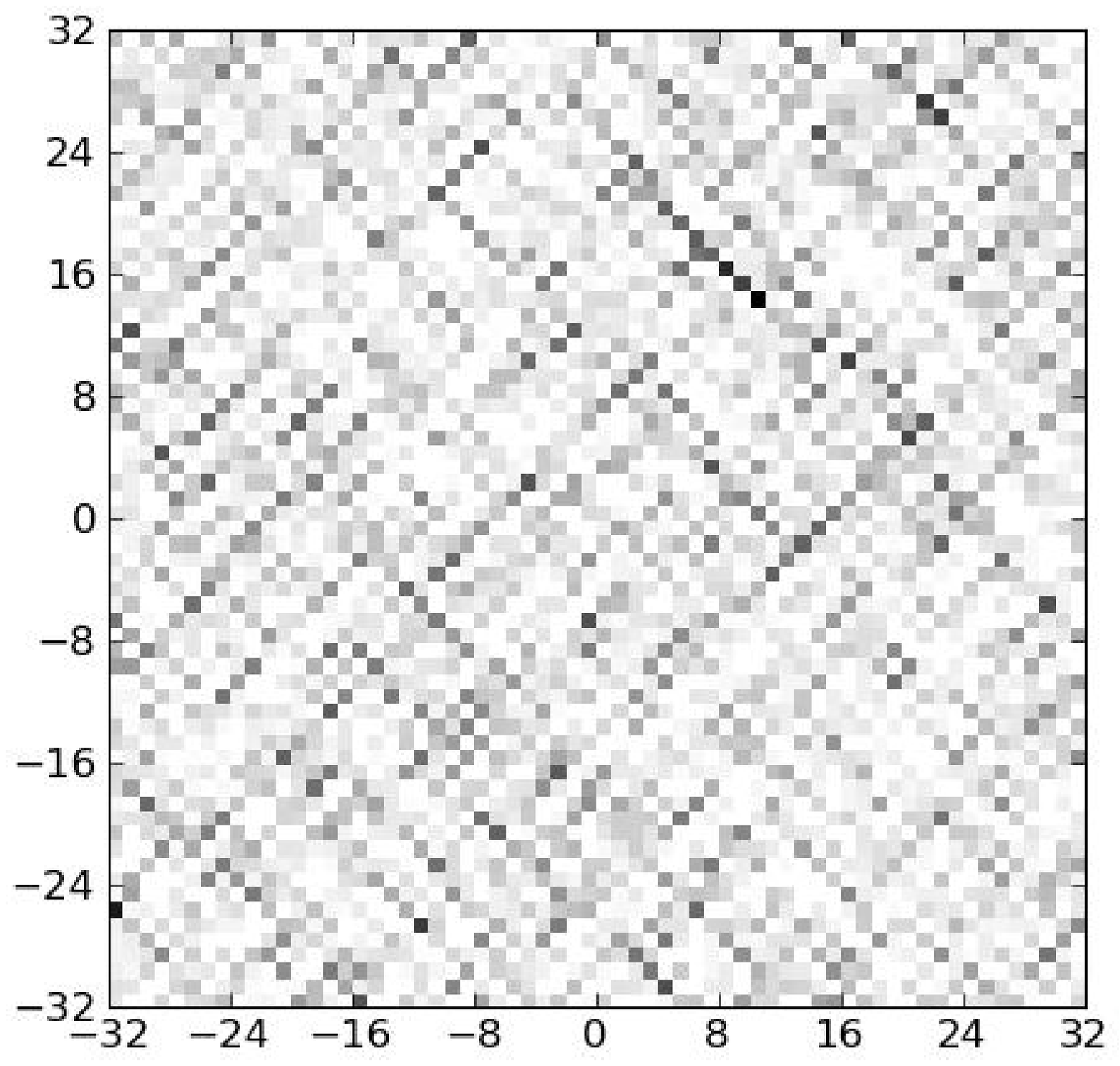}
\includegraphics[width=0.195\textwidth]{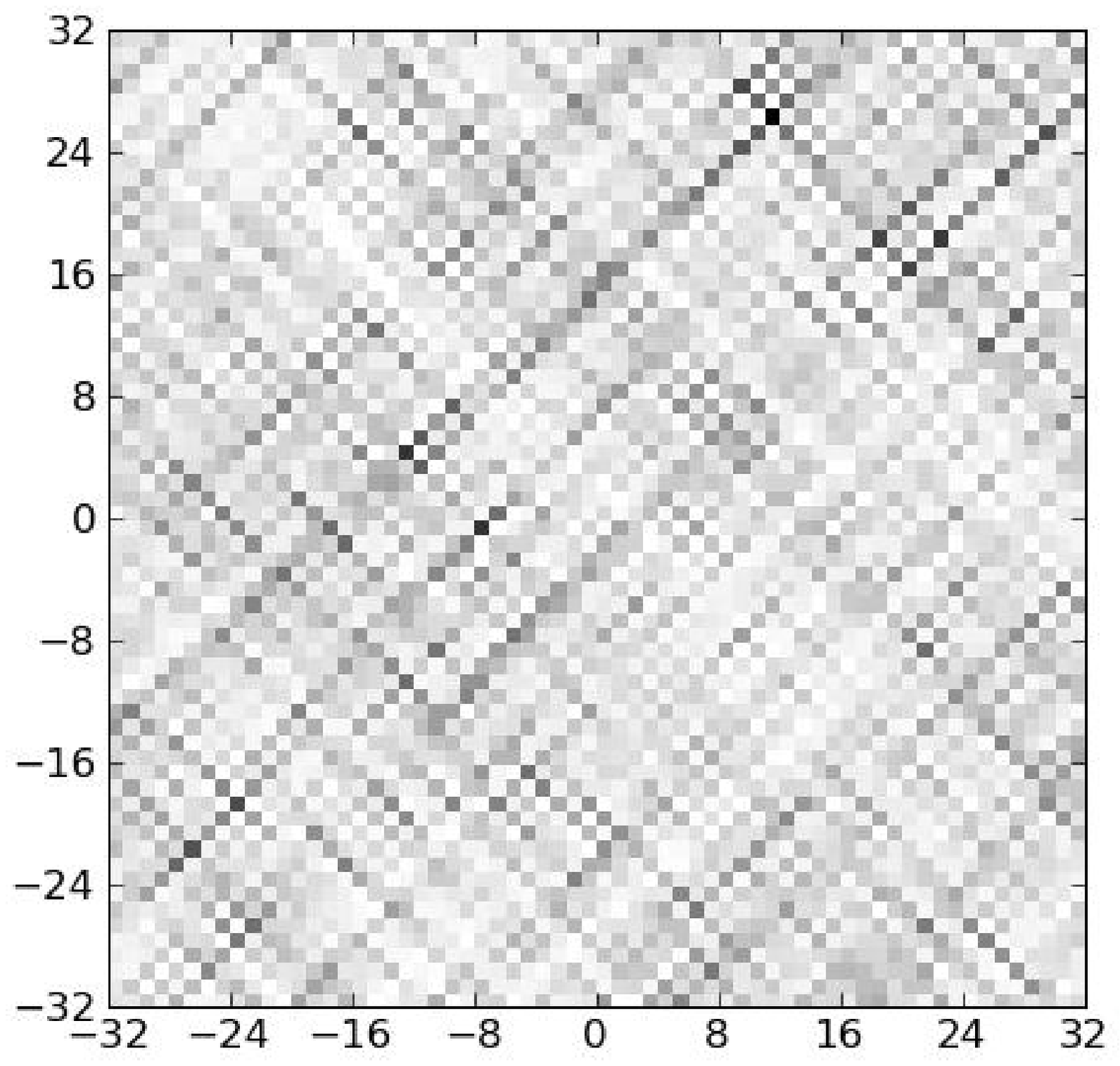}
\includegraphics[width=0.195\textwidth]{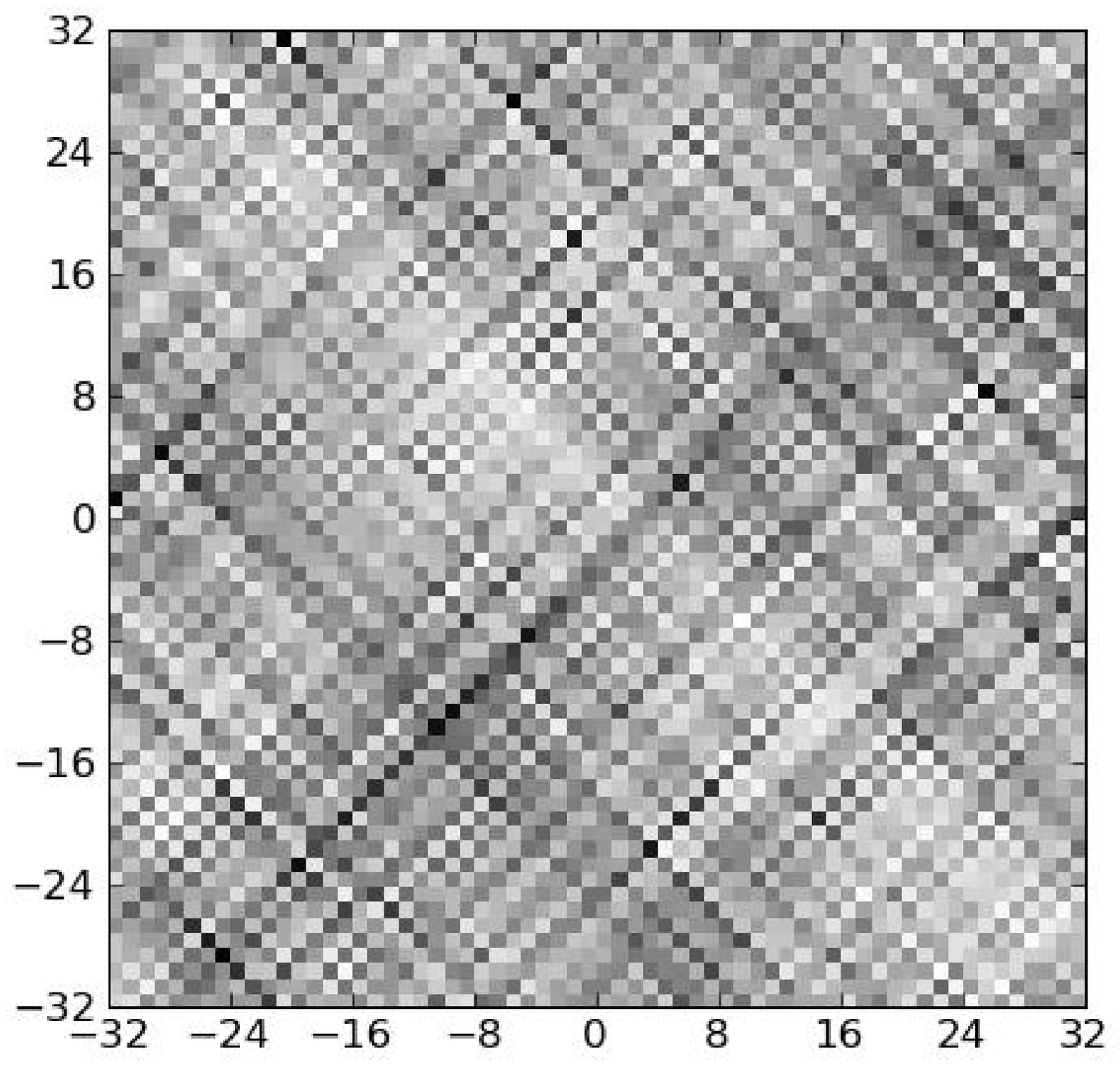}
\includegraphics[width=0.195\textwidth]{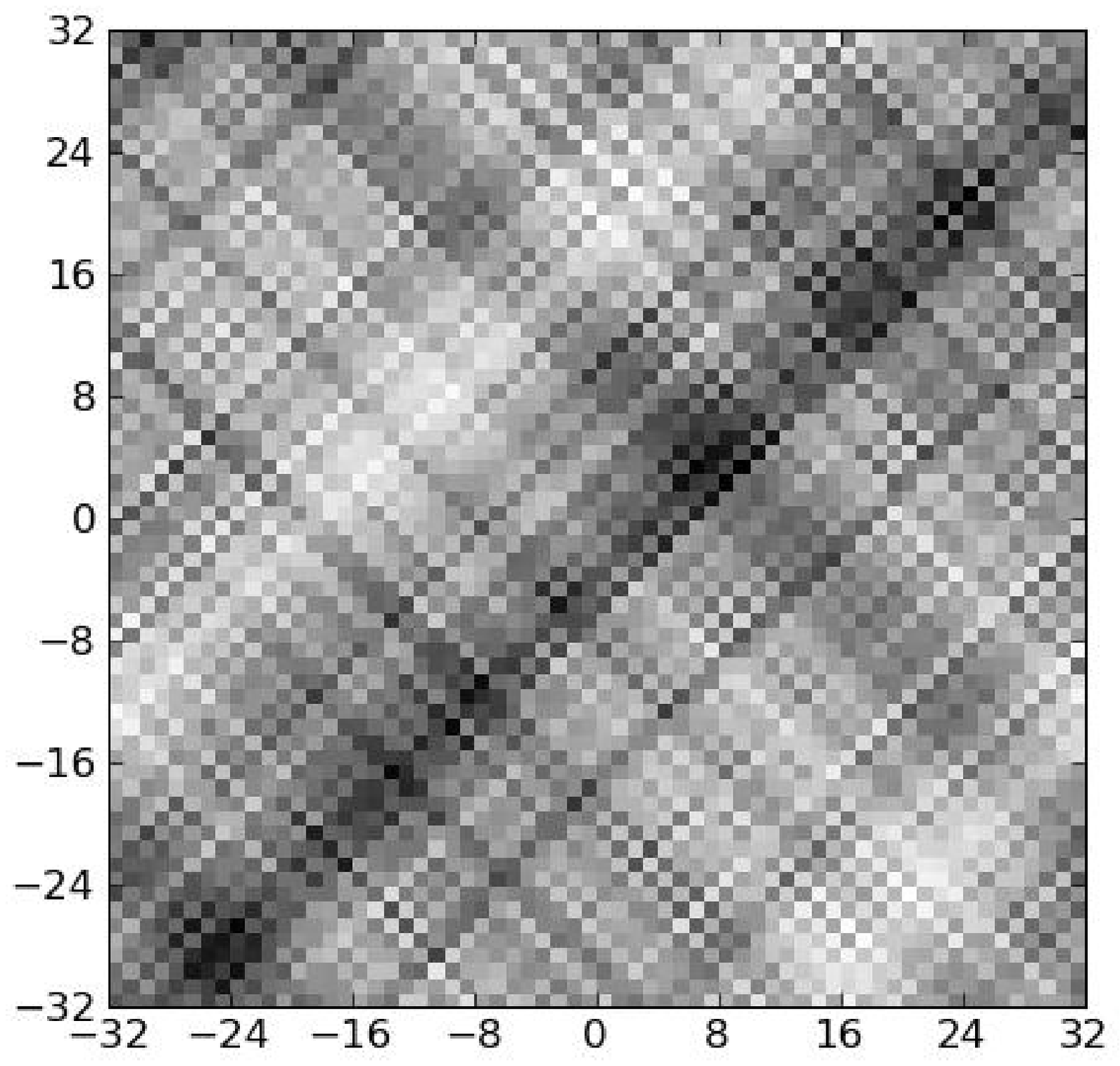}
 \end{minipage}
\begin{minipage}[h]{1.0\textwidth}
\includegraphics[width=0.195\textwidth]{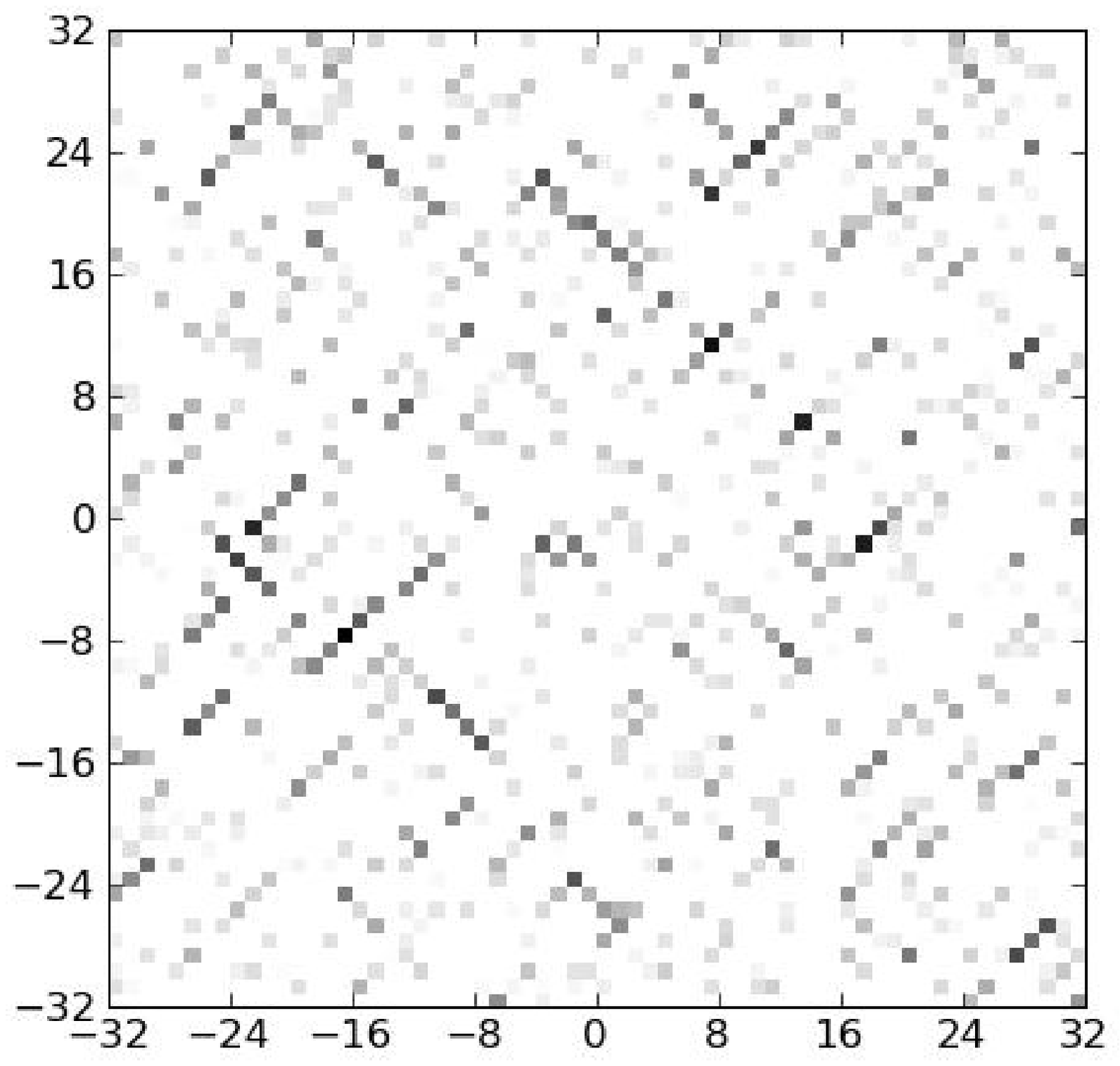}
\includegraphics[width=0.195\textwidth]{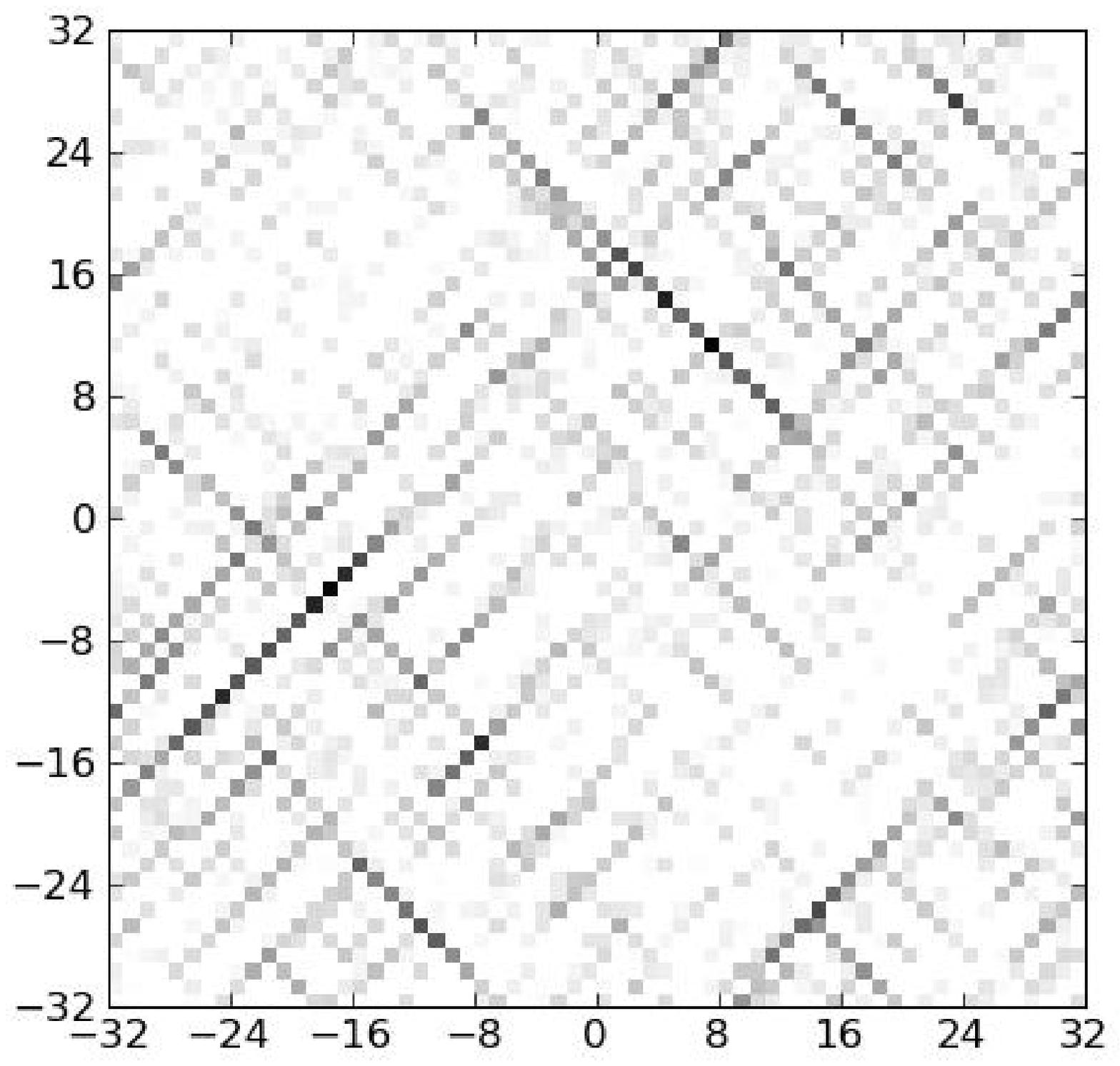}
\includegraphics[width=0.195\textwidth]{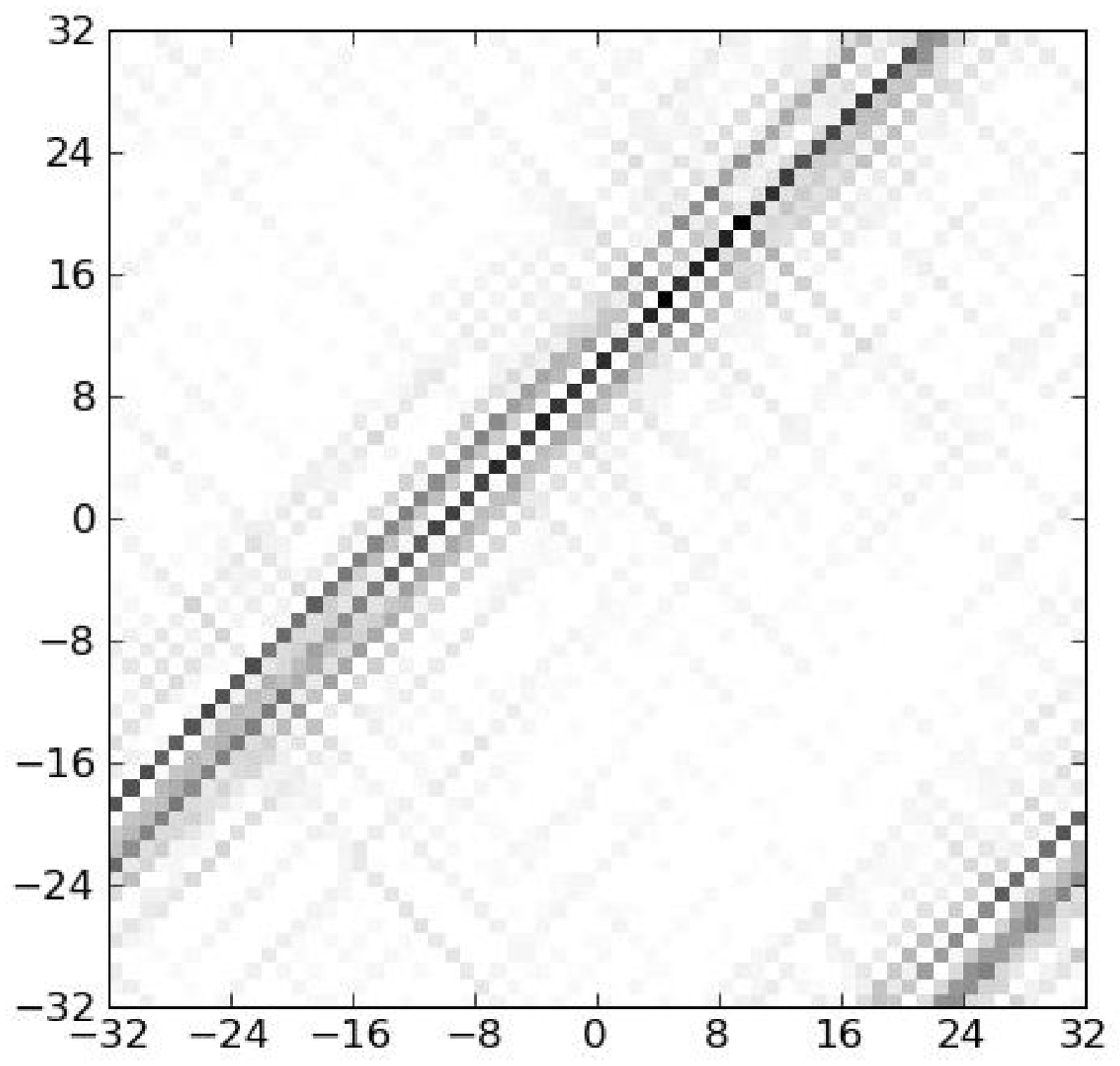}
\includegraphics[width=0.195\textwidth]{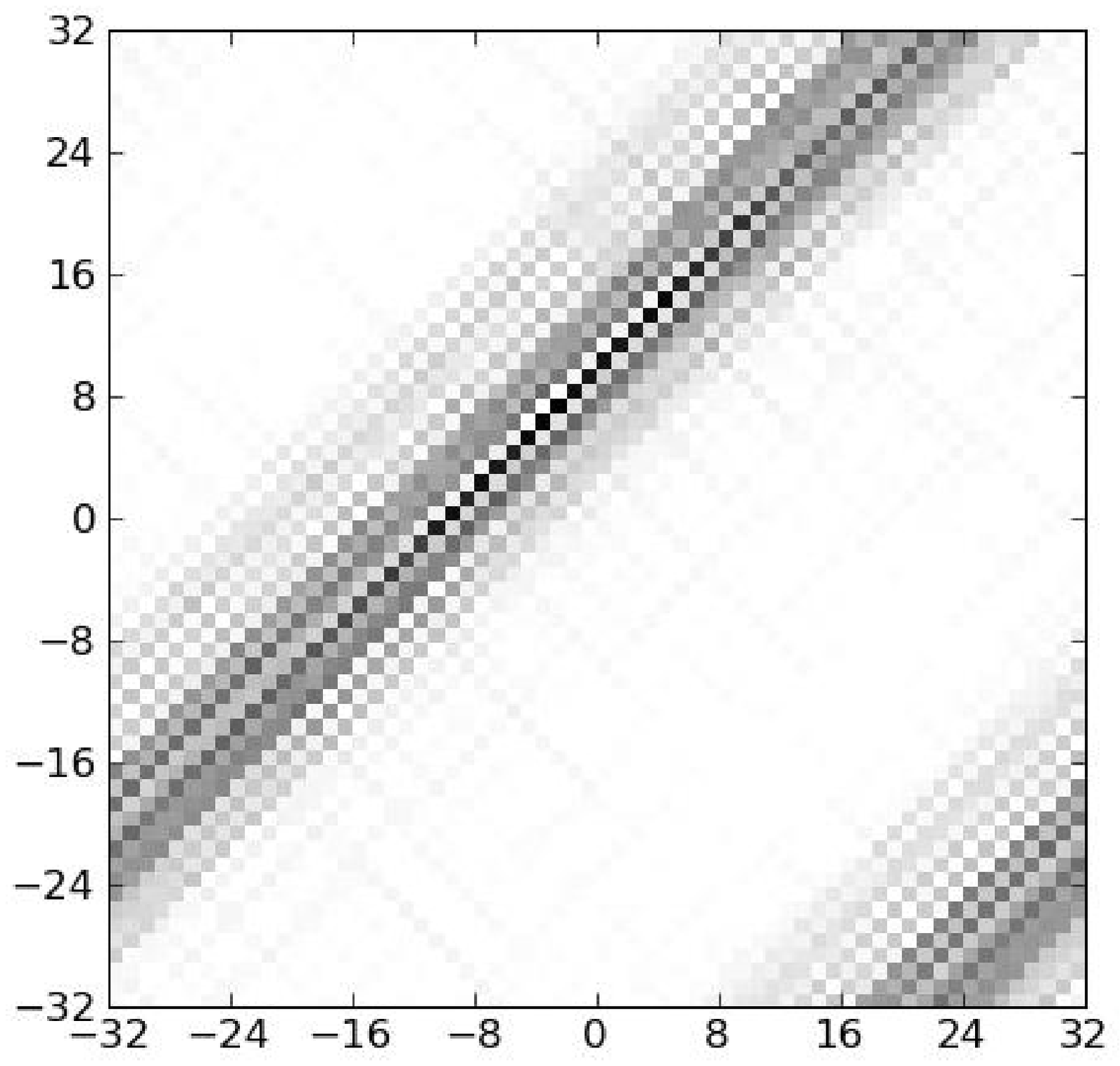}
\includegraphics[width=0.195\textwidth]{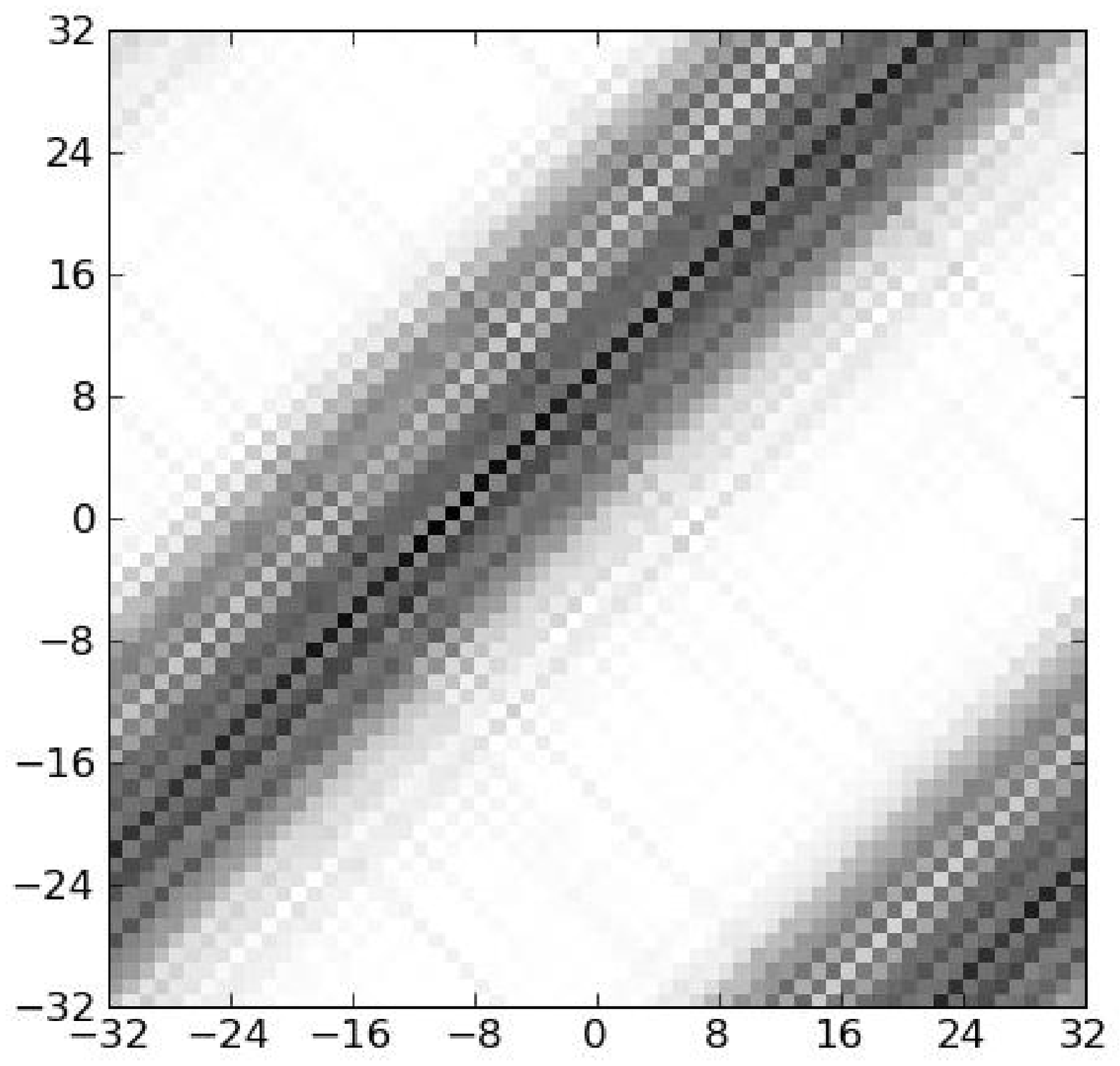}
 \end{minipage}
\begin{minipage}[h]{1.0\textwidth}
\includegraphics[width=0.195\textwidth]{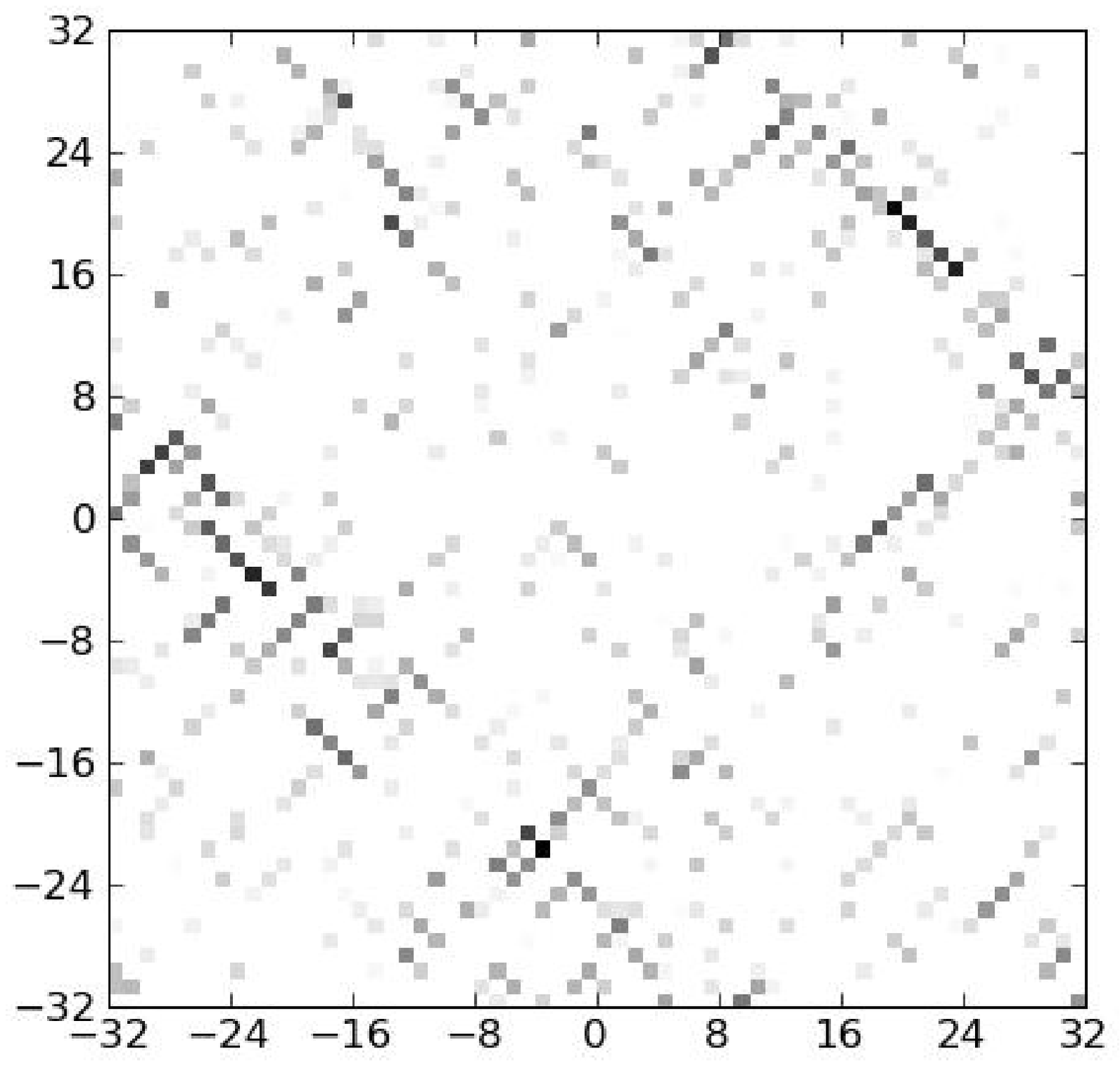}
\includegraphics[width=0.195\textwidth]{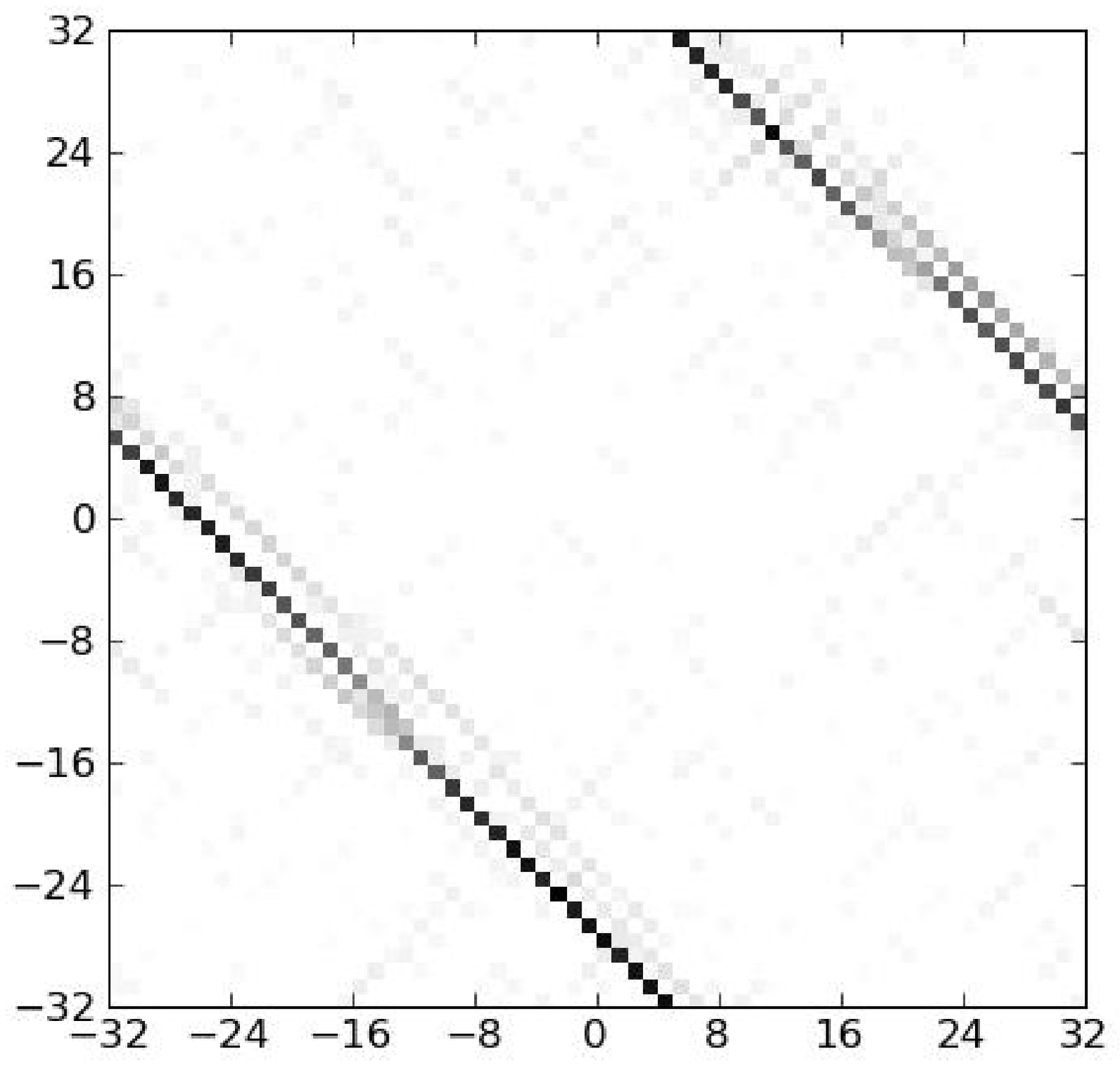}
\includegraphics[width=0.195\textwidth]{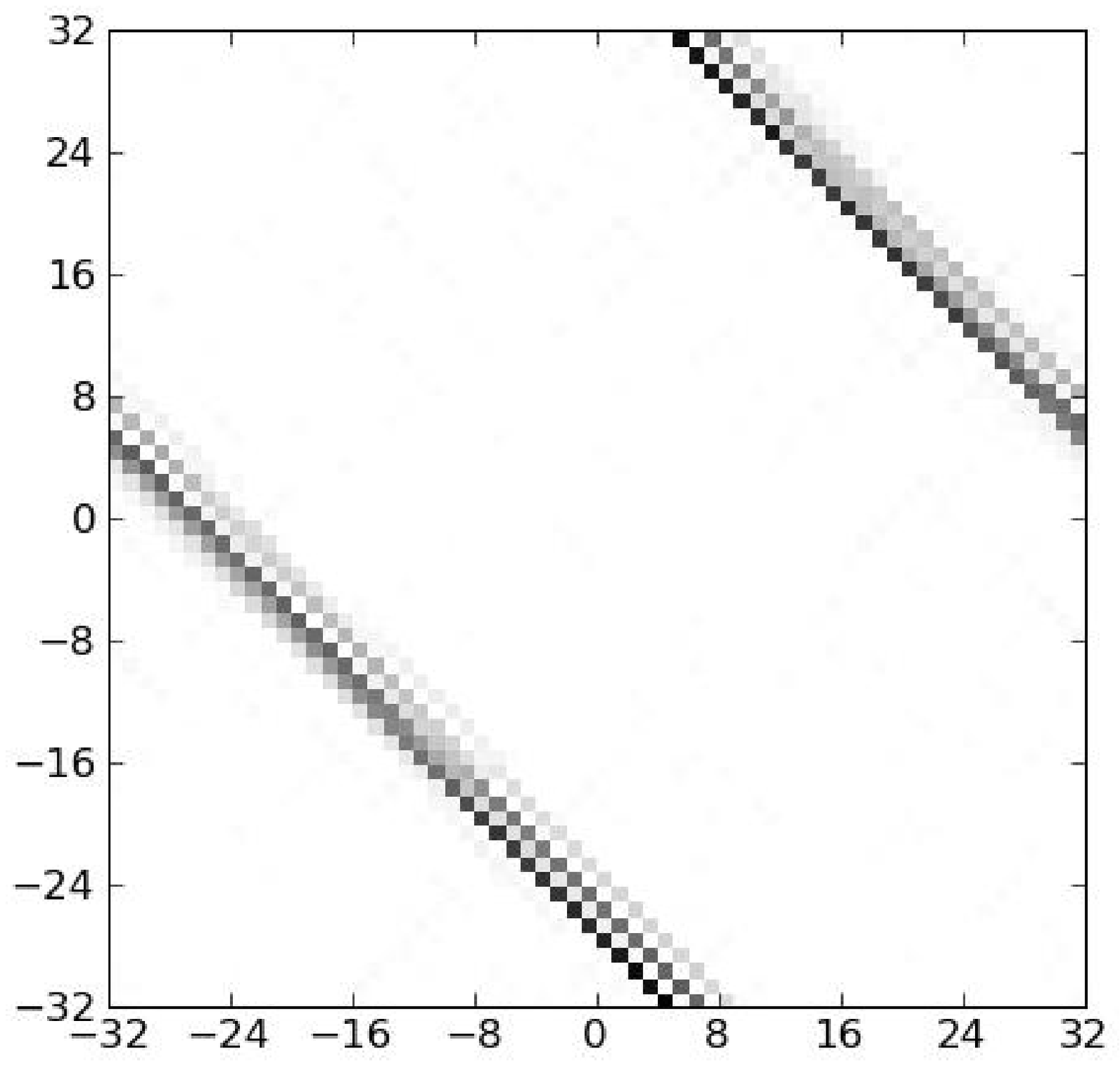}
\includegraphics[width=0.195\textwidth]{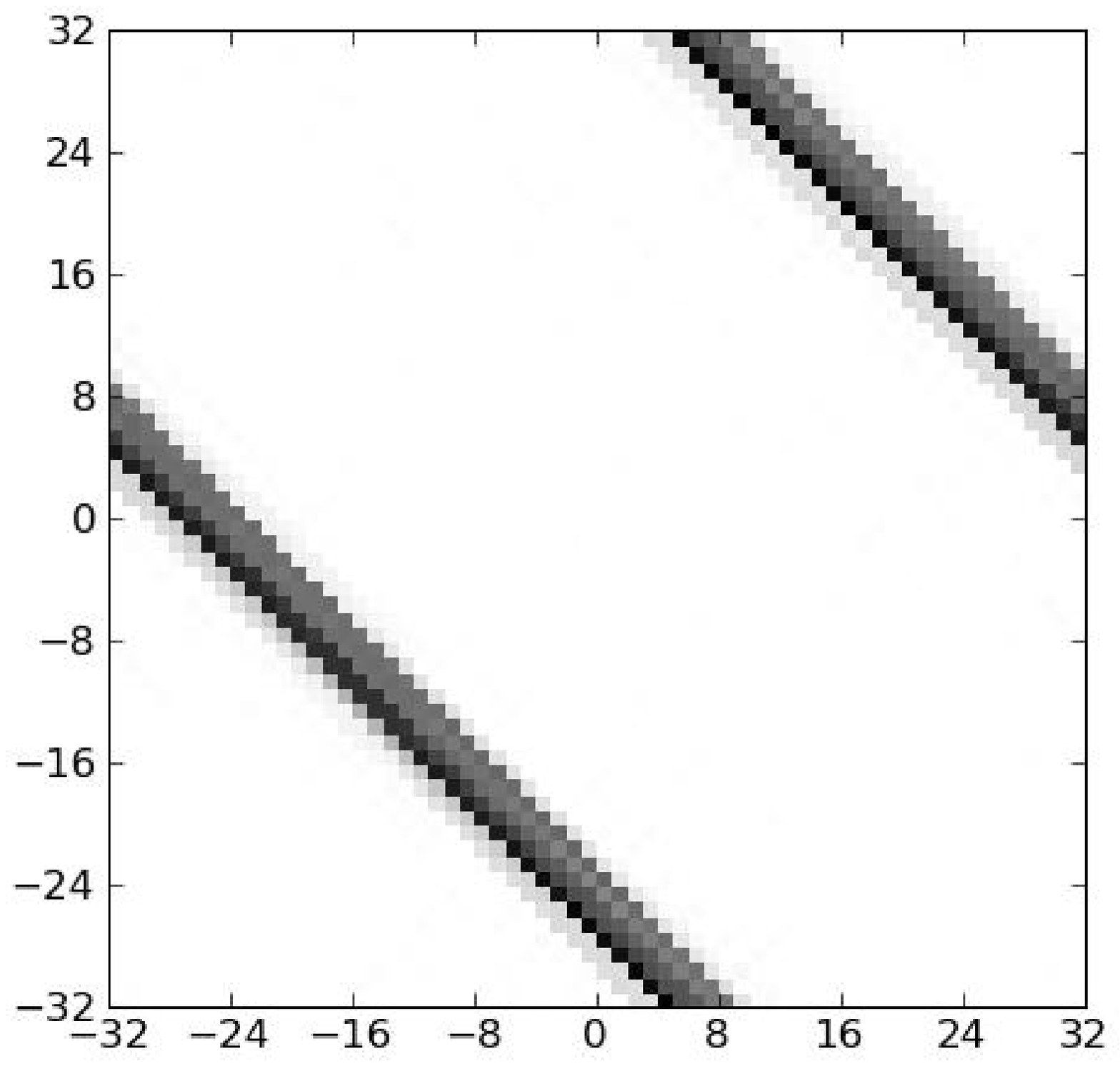}
\includegraphics[width=0.195\textwidth]{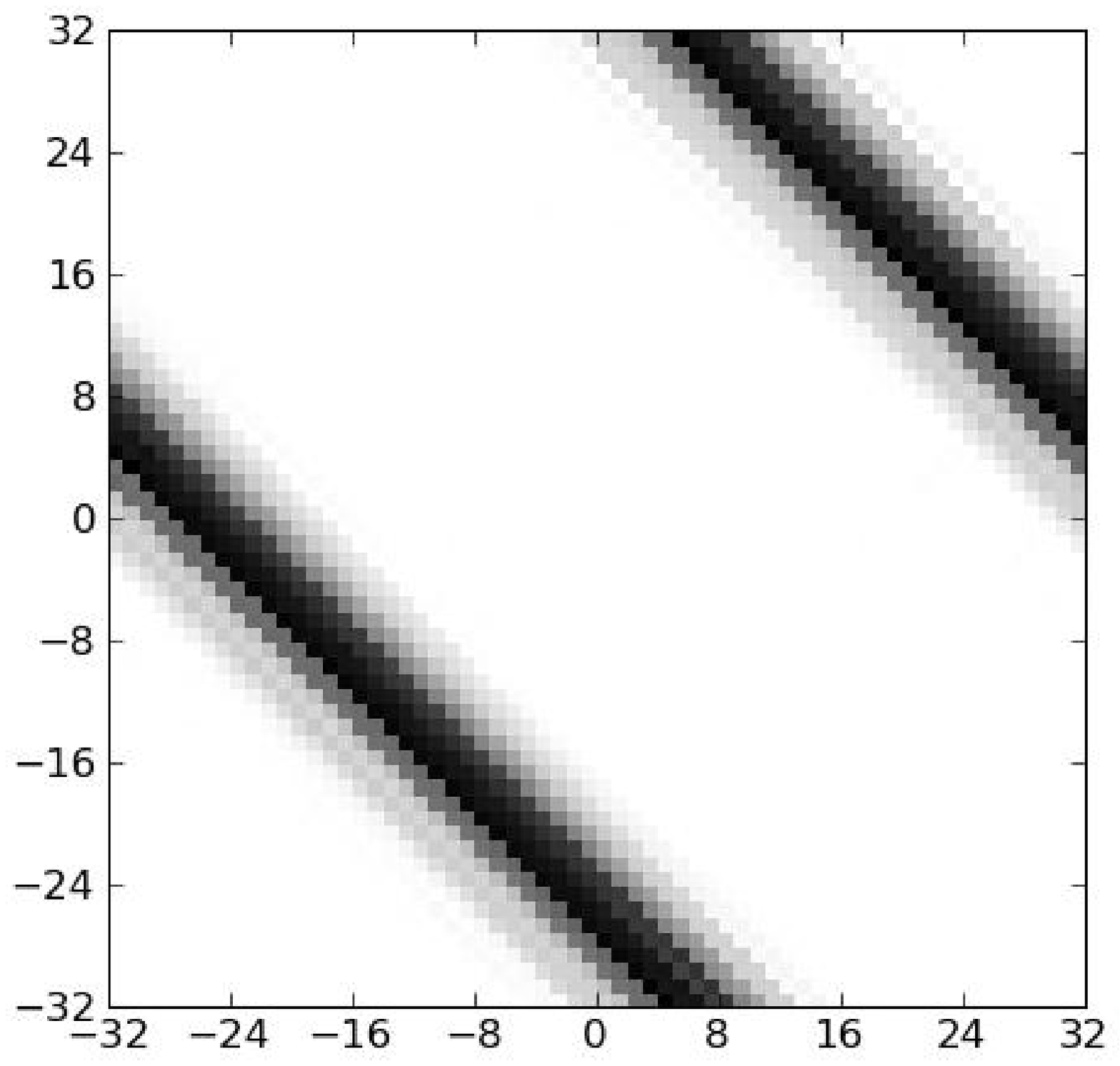}
 \end{minipage}
\caption{Maps of plastic strain obtained from left to right at
$\langle\varepsilon_p\rangle = 1/16$, $1/4$, $1$, $4$ and $16$ and
from top to bottom with a bias value $\delta = 0$, $0.5$ and
$0.7$  with a slip increment $d=0.3$.}
\label{ActivityMaps-vs-delta}
\end{figure*}

\section{Definition of the model}

Let us briefly recall the definition of the model (see
Ref. \cite{TPRV-meso10} for more details).
 The mechanical fields are discretized on a square lattice with a
mesh size significantly larger than the typical scale of a plastic
reorganization.  Periodic boundary conditions are considered. The
material is assumed to be elastically homogeneous, so that stresses
and elastic moduli are scaled so that the steady-state local yield
stress is unity. A local criterion of plasticity is considered. The
initial distribution of local yield stress is denoted
$P_i(\sigma_c)$. Every time a local plastic criterion is satisfied at
point $\bm x_0$, a local slip $ \Delta\varepsilon_p$ occurs (we assume
here that local plastic strains obey the same symmetry as the external
loading, pure shear in the present case, so that a simple scalar yield
criterion can be chosen) with a random amplitude $d$ drawn from a
statistical distribution $Q(d)$, $ \Delta\varepsilon_p(\bm x)=d
\delta_D(\bm x-\bm x_0)$ where $\delta_D$ is the Dirac distribution. Note
that $d$ is the product of the mean plastic strain by the ``volume''
of the transformation zone. This local slip $d$ induces a long range
redistribution of elastic stress with a quadrupolar symmetry (see Ref.
\cite{TPRV-meso10,TPRV-PRE08} for analytical and numerical details
about this elastic propagator) $\Delta\sigma_{el}(\bm x)=d G(\bm x-\bm
x_0)$ with $G(r,\theta) \approx Ad\cos4\theta/r^ 2$ where $A$ is the
dimensionless elastic constant, $r$ and $\theta$ the polar
coordinates. The slip amplitude $d$,
is drawn from a uniform distribution in the range $[0;d_0]$.

After slip, the microstructure of the flipping zone has changed and a
new value of the local yield stress is drawn from a distribution
$P_S(\sigma_c)$. The system is driven with an extremal dynamics so
that only one site at a time is experiencing slip.  The originality of
the present depinning models relies in the anisotropic elastic
interaction.  Within this framework of dynamic phase transition, the
choice of extremal dynamics ensures to drive the system at the verge
of criticality: the macroscopic yield stress is given by the critical
threshold.

The above model may be seen as a depinning model for amorphous
plasticity with a peculiar (anisotropic) elastic interaction.  While
the richness of the physics of depinning models mainly relies on the
competition between elasticity and disorder, we see here that the
anisotropic character and the abundance of soft modes in the elastic
interaction which characterize the present model of amorphous
plasticity, naturally induce an additional competition between
localization and disorder.

An implicit assumption used in our model is that the
  statistical distribution $P_S(\sigma_c)$ used to renew the local
  plastic threshold under shear ({\it i.e.}  after local slip) is the
  very same as the distribution of plastic thresholds in the initial
  configuration $P_i(\sigma_c)$. This hypothesis may
  be questioned.  Indeed, various experimental and numerical results
  obtained in friction or in shearing granular material or complex
  fluids~\cite{Baumberger-AdvPhys06,Falk-PRL07,Behringer-GM10} seem to
  indicate an effect of the preparation of the material upon its behavior
  under shear. One may think for instance at the effect of density of
  granular material: a loose (dense) packing tends to exhibit
  hardening (softening) while under shear their density progressively
  evolves toward a so-called ``critical'' value.

In order to test the effect of our hypothesis we give in the following
a bias to the initial thresholds distribution and try to test its
consequences. Practically speaking, the yield stress distributions
$P_i(\sigma_c)$ (initial state) and $P_S(\sigma_c)$ (under shear) are
chosen as uniform in the ranges $[\delta;1+\delta]$ and $[0;1]$
respectively.  A positive (negative) value of $\delta$ is expected to
induce some softening (hardening) behavior since all threshold values
above unity (below zero) should eventually be replaced by thresholds
within the interval $[0,1]$.

We focus in the following discussion on the effect of
these two parameters $d_0$ and $\delta$.  In the view developed by
Sollich et al. \cite{Sollich-PRL97}, the parameter $d_0$ which gives
the amplitude of the mechanical noise induced by the elastic
interactions may be thought as analogous to the effective mechanical
temperature $x$ in the SGR model.  However, it is to be emphasized
that this mechanical ``noise'' is strongly inhomogeneous in space and
displays strong temporal and spatial correlations, absent from the SGR
model. The second parameter $\delta$ measures the shift between the
initial yield stress distribution, that  uniform in $[\delta;1+\delta]$ and
the distribution of new local yield stress under shear, uniform in the
range $[0;1]$ may be related to the initial state of the system prior
to shearing.  Indeed it is expected that the older the glassy system,
the more stable and the more difficult it is to shear.  This effect is
described here through a mere penalty in the initial yield
stress. High mean values of the plastic thresholds should thus be
associated with aged configurations of the glass. As discussed in
\cite{Rottler-PRL05}, a logarithmic increase of the yield stress with
the age of the system is often observed in glassy materials: $\delta
\approx s_0(T) \log(t_w/t_0)$ where $t_0$ is a microscopic time
scale. According to this perspective, the age of the system would
simply be related to the bias $\delta$ through an exponential
dependence.
Yet another interpretation would consist of relating
  the parameter $\delta$ to a structural temperature as discussed in
  Ref.\cite{Falk-PRL07,Manning-PRE07,Manning-PRE09,Bouchbinder-PRE09b}. The two
  quantities are expected to vary in opposite ways when the structure
  relaxes. The more relaxed the glass, the higher $\delta$ and the
  lower the structural temperature. We now test these simple ideas
against numerical simulations.

\section{Maps of plastic activity}

\begin{figure*}[t]
\begin{minipage}[h]{1.0\textwidth}
\includegraphics[width=0.195\textwidth]{strain005-05-2.eps}
\includegraphics[width=0.195\textwidth]{strain005-05-4.eps}
\includegraphics[width=0.195\textwidth]{strain005-05-6.eps}
\includegraphics[width=0.195\textwidth]{strain005-05-8.eps}
\includegraphics[width=0.195\textwidth]{strain005-05-10.eps}
 \end{minipage}
\begin{minipage}[h]{1.0\textwidth}
\includegraphics[width=0.195\textwidth]{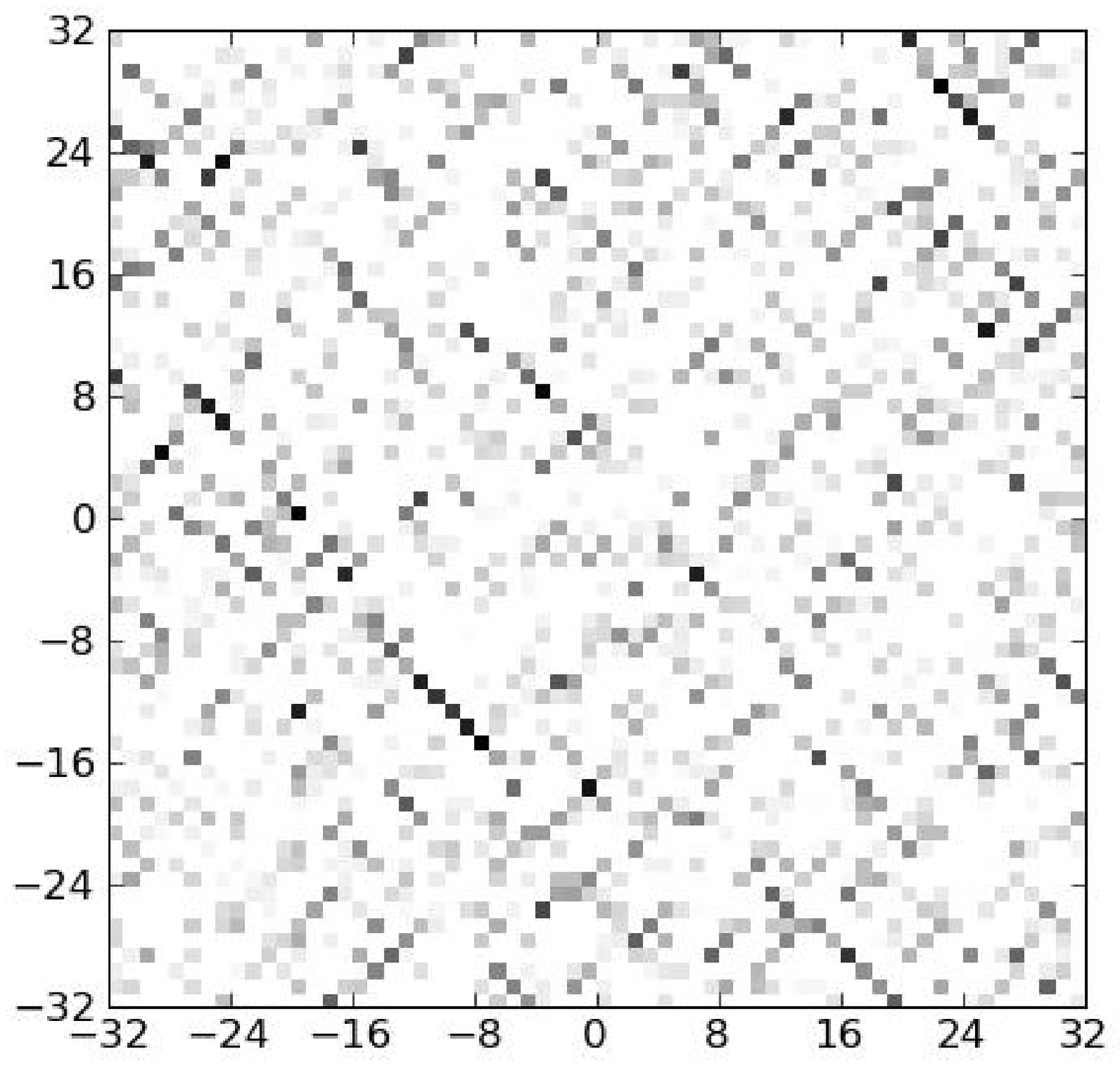}
\includegraphics[width=0.195\textwidth]{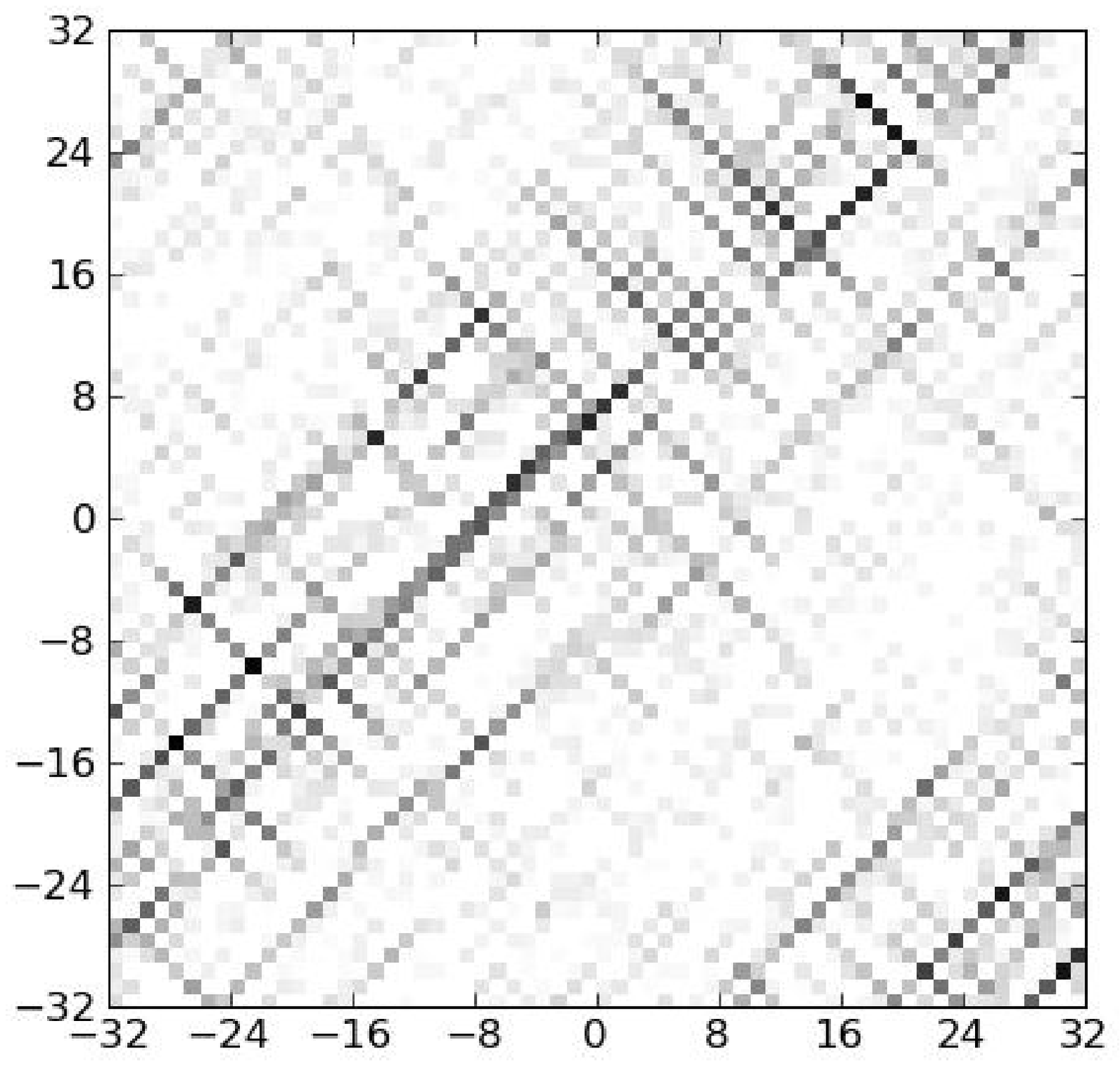}
\includegraphics[width=0.195\textwidth]{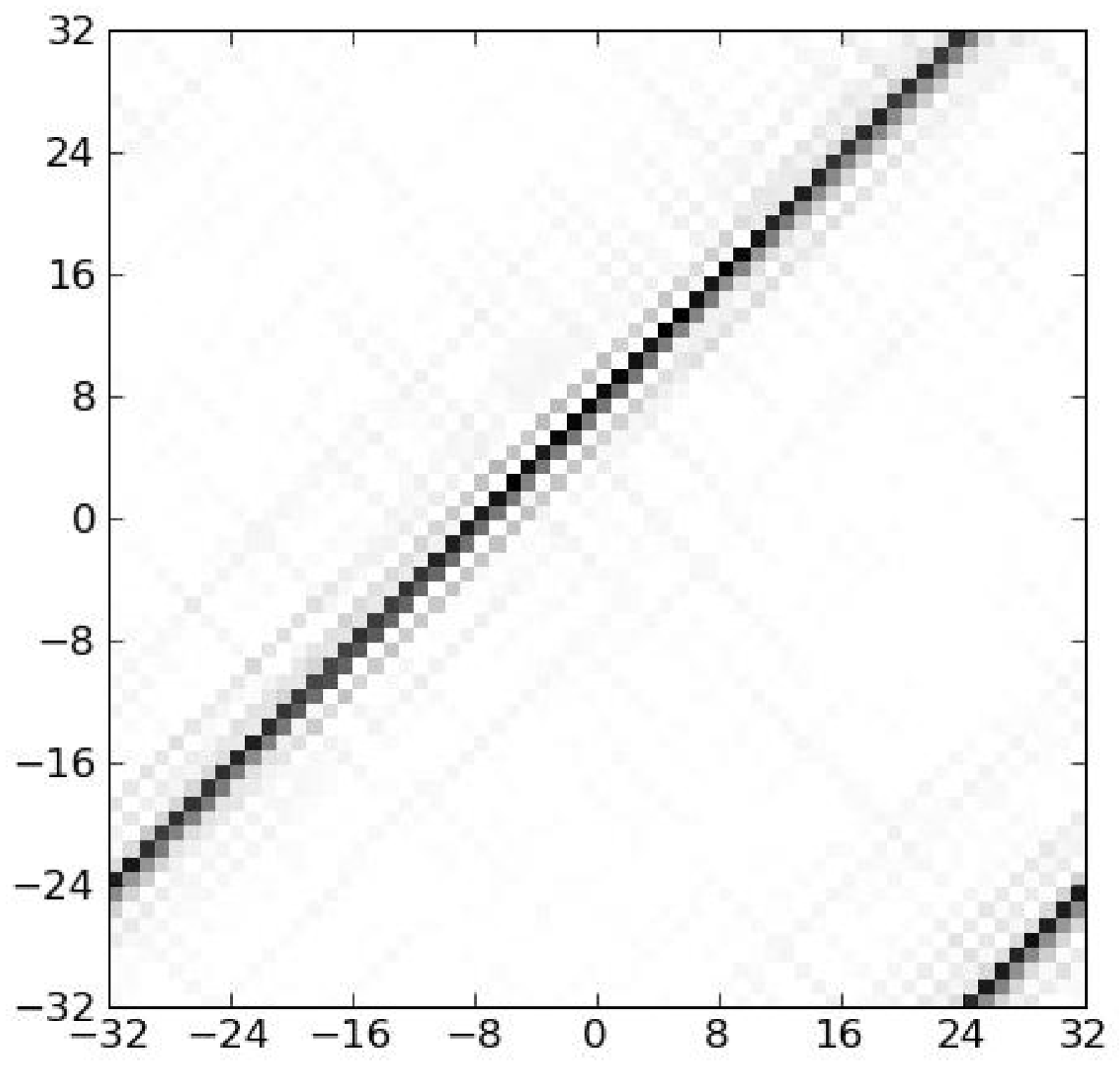}
\includegraphics[width=0.195\textwidth]{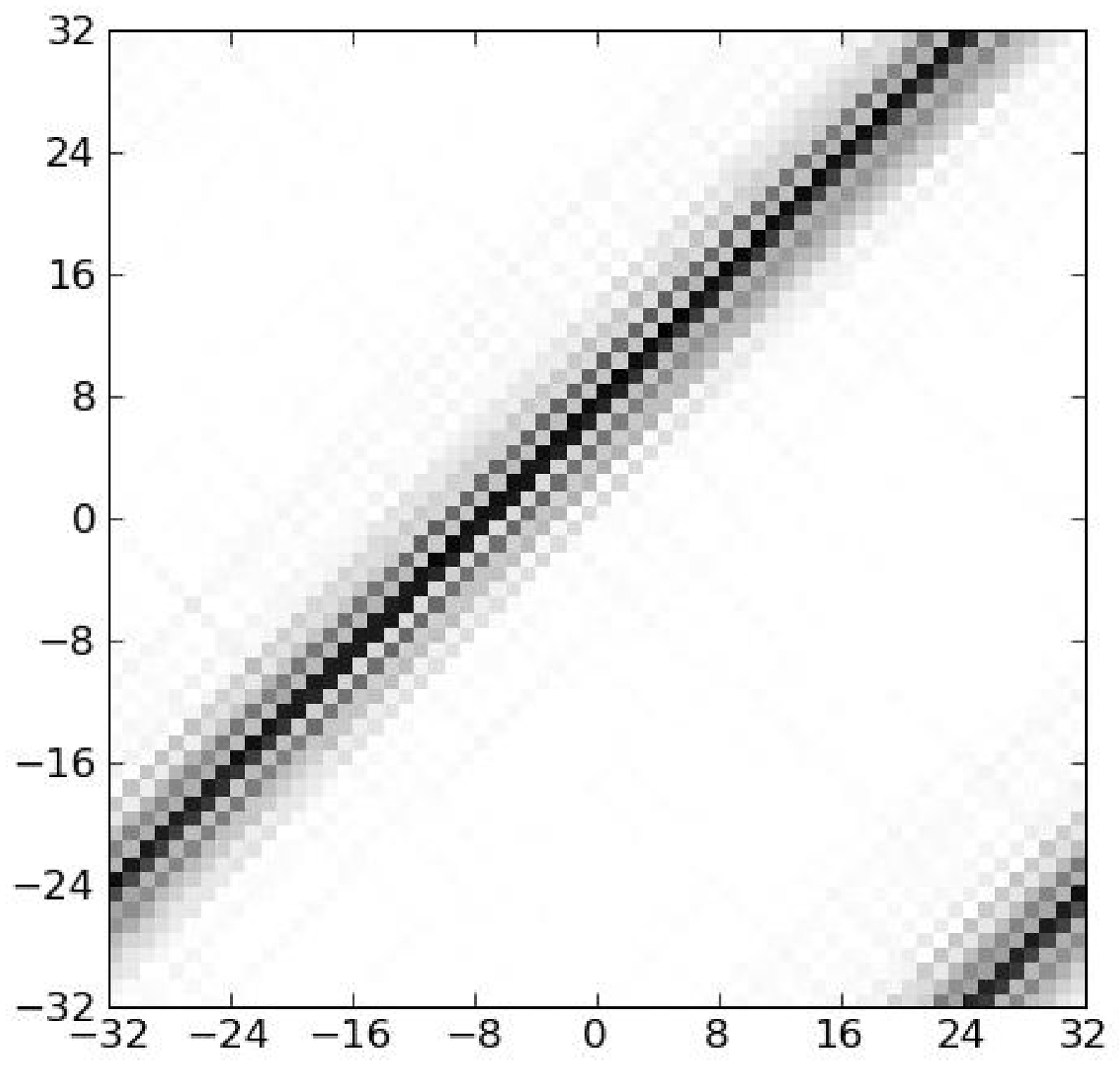}
\includegraphics[width=0.195\textwidth]{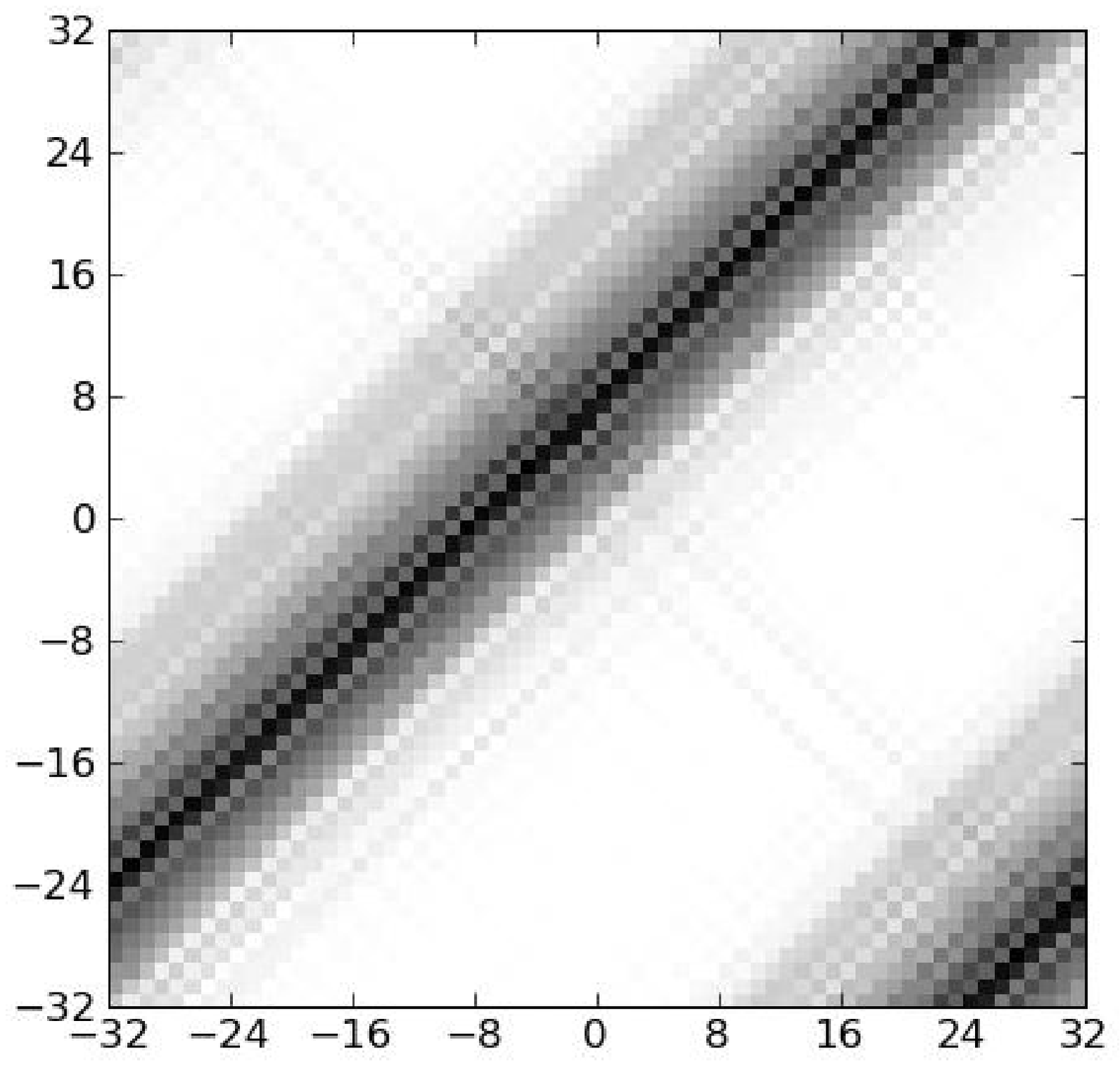}
 \end{minipage}
\begin{minipage}[h]{1.0\textwidth}
\includegraphics[width=0.195\textwidth]{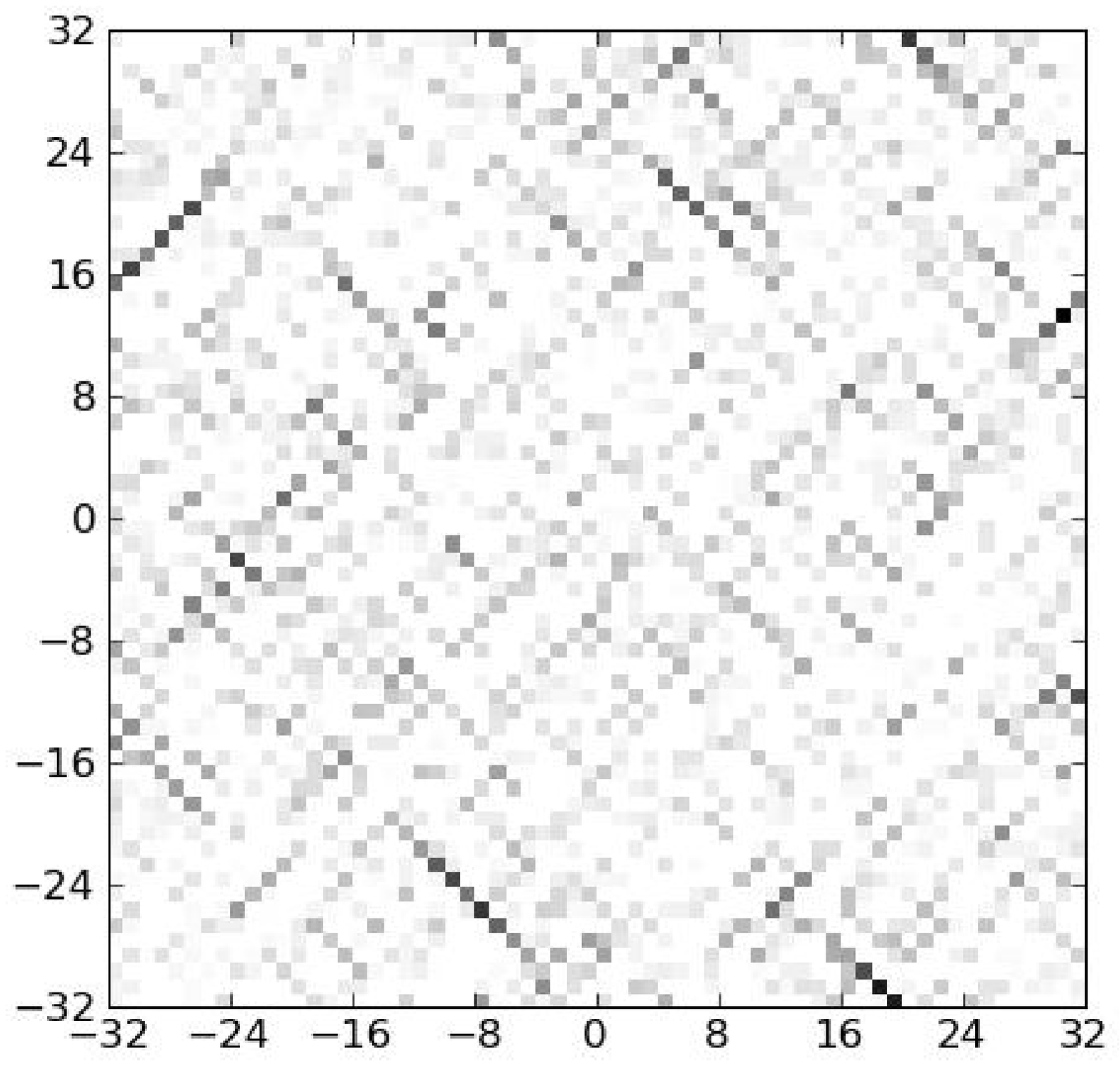}
\includegraphics[width=0.195\textwidth]{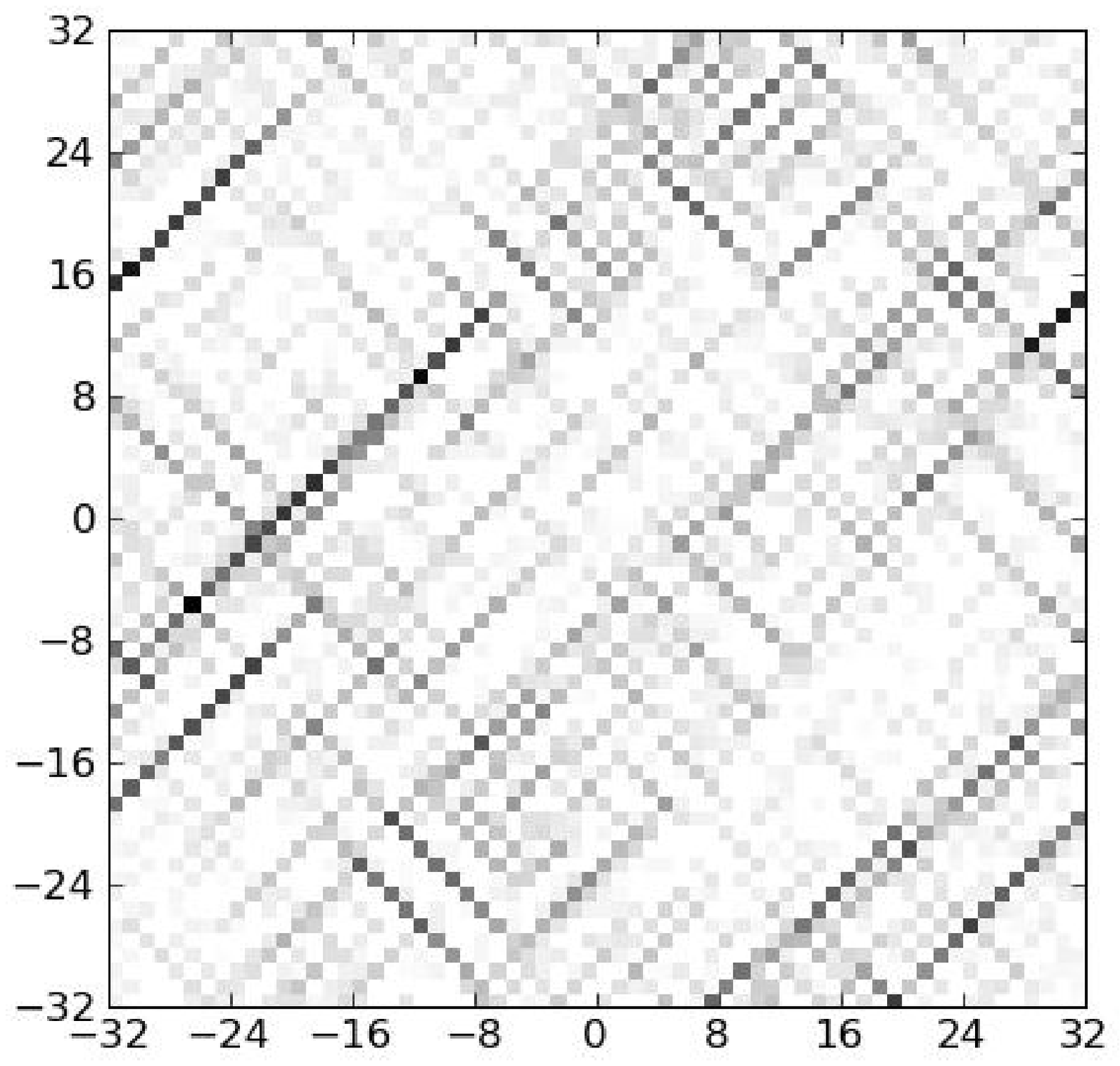}
\includegraphics[width=0.195\textwidth]{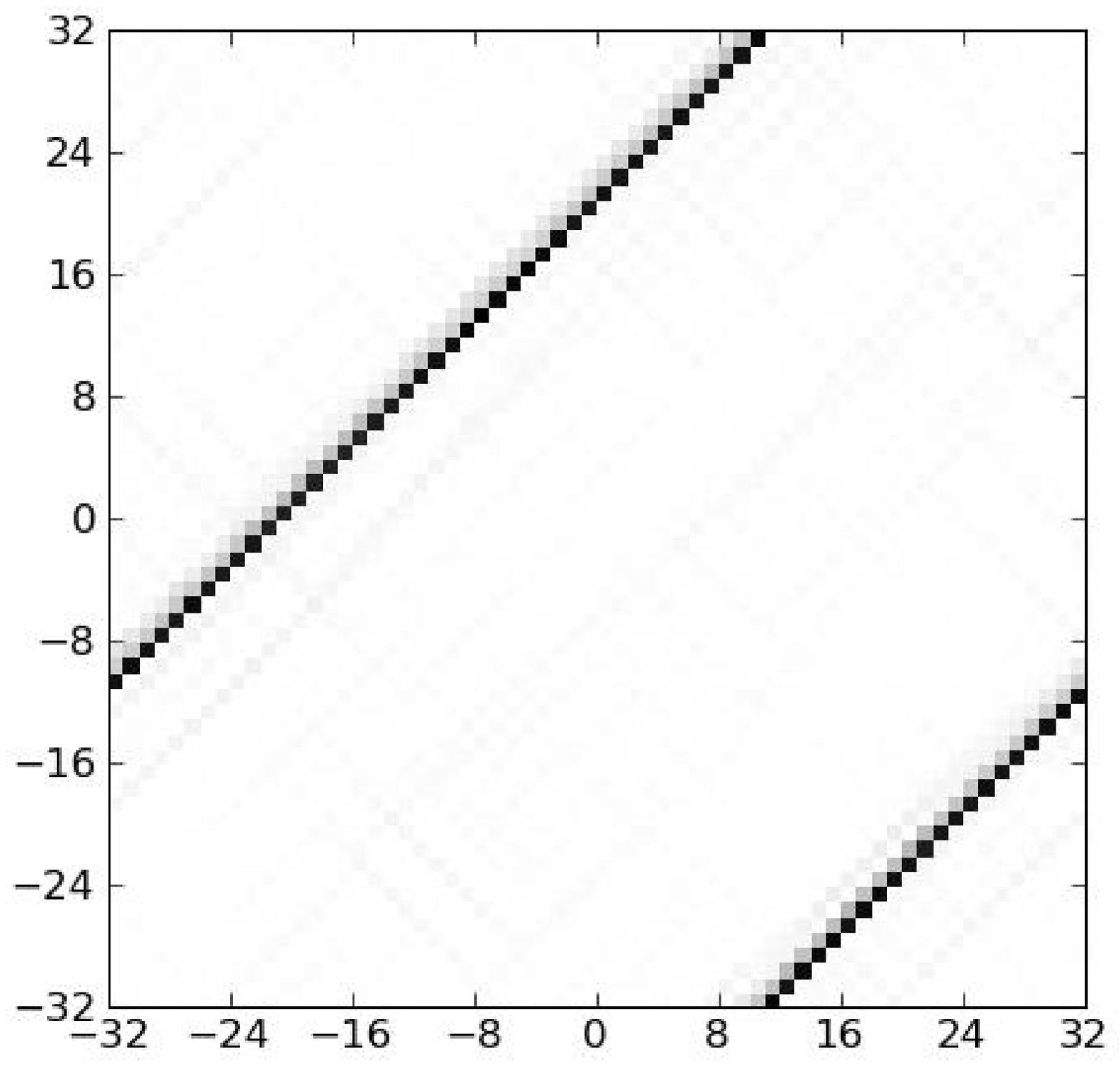}
\includegraphics[width=0.195\textwidth]{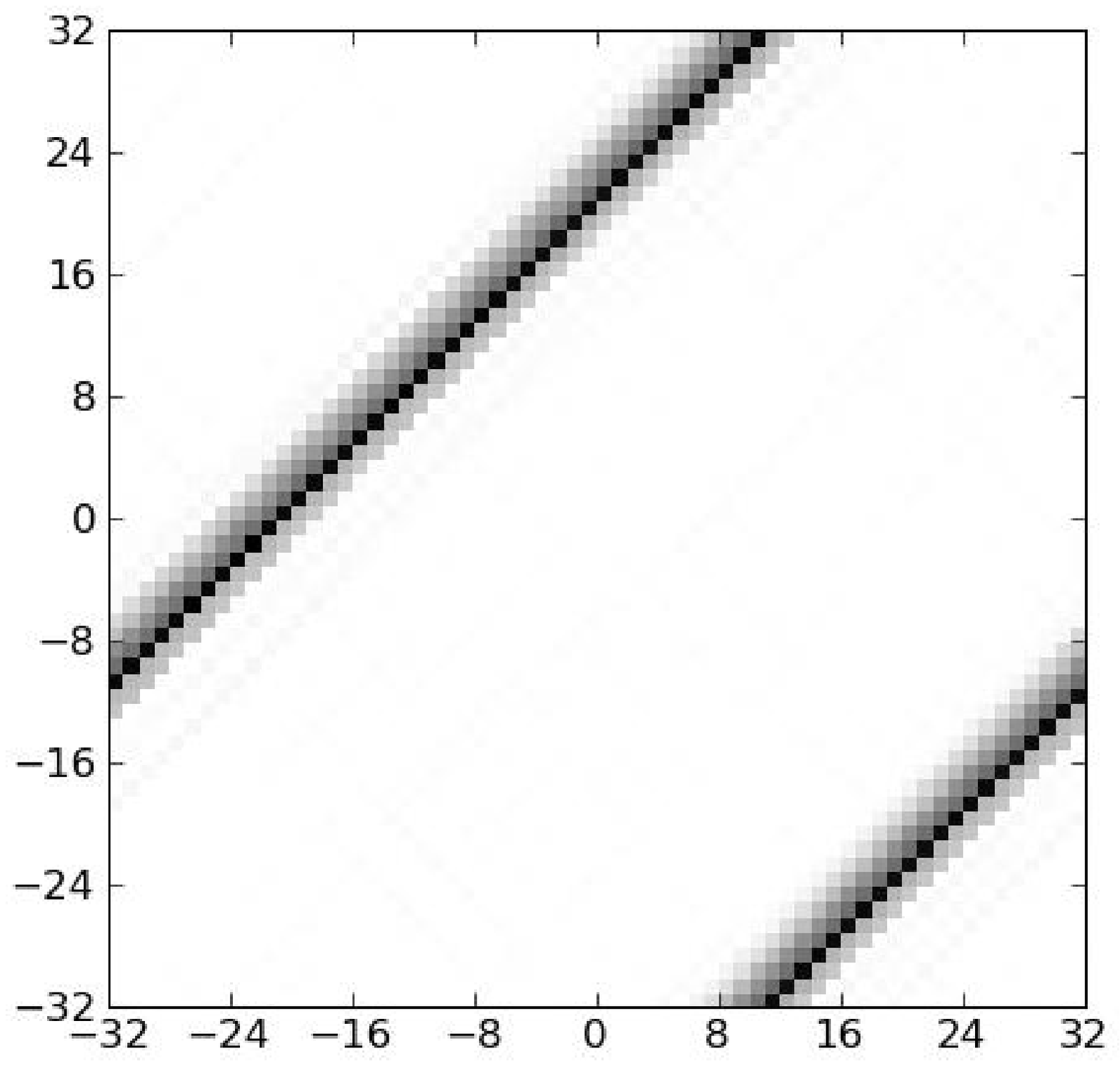}
\includegraphics[width=0.195\textwidth]{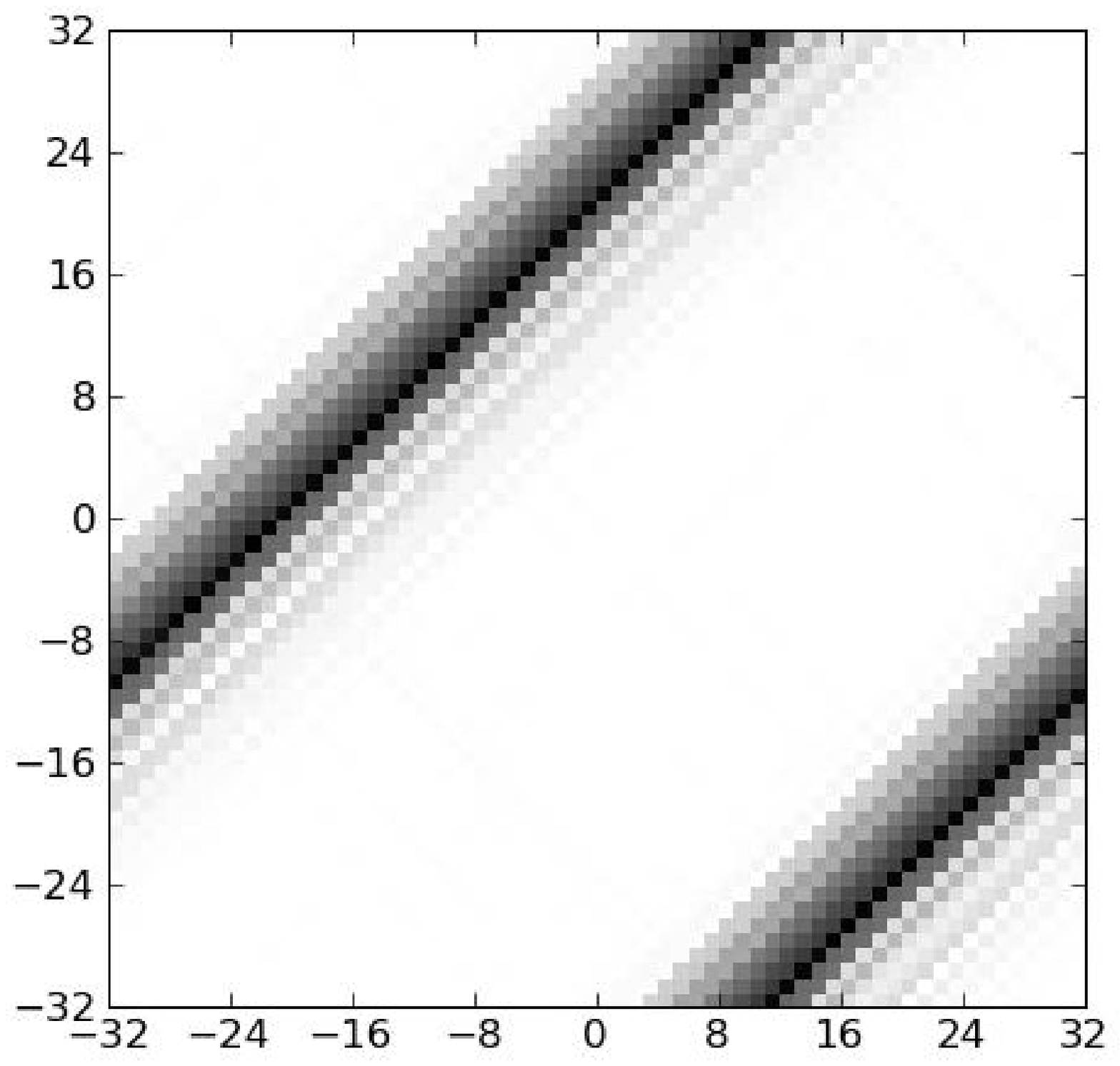}
 \end{minipage}
\caption{Maps of plastic strain obtained from left to right at
$\langle\varepsilon_p\rangle = 1/16$, $1/4$, $1$, $4$ and $16$ with a bias value
$\delta = 0.5$, and from top to bottom with a slip increment $d_0 =0.3$, $0.1$
and $0.03$. } \label{ActivityMaps-vs-d0}
\end{figure*}

In Fig. \ref{ActivityMaps-vs-delta} and
\ref{ActivityMaps-vs-d0} the evolution of the spatial distribution of plastic
strain under shear is shown for different values of the age-like parameter
$\delta$ (Fig. \ref{ActivityMaps-vs-delta}) and of the temperature-like
parameter $d_0$. We show snapshots of the plastic strain field taken at $\langle
\varepsilon_p \rangle=1/16,\;1/4,\;1,\;4,\;16$. The value of the local strain is
represented with a grey scale, (the darker, the larger plastic strain).

In Fig. \ref{ActivityMaps-vs-delta}, the values $d_0=0.3$ of the slip increment
has been used. The first row corresponds to the value $\delta = 0$. When using
this un-aged initial configuration, we see that plastic strain first
self-organizes along shear bands at $\pm \pi/4$ {\it i.e.,} according to the
maximum shear directions. This localization is however not persistent and after
a transient, these shear bands diffuse throughout the system.  The evolution
obtained with a bias value $\delta=0.5$ (second row) is markedly different.
Again plastic deformation first tends to form shear bands according to
directions at $\pm \pi/4$, but remains essentially trapped in a strongly
localized state. The formed shear band only slowly widens with ``time'' (mean
plastic strain). The evolution obtained with a bias value $\delta=0.7$ (third
row) is very similar : formation of a persistent shear band before an apparent
diffusive widening of the band. Localization appears to be more intense and
widening slower with this higher value of the age-like parameter $\delta$.

Let us only note here that the way the plastic activity gets localized along a
band is somewhat reminiscent of the behavior of an earlier model proposed by
Torok and Roux \cite{Torok-PRL00}. In this study, the authors made evidence for
a weak breaking of ergodicity which they relate to the progressive building in
the thresholds landscape of a valley (along the shear band) surrounded by ridges
elevating significantly above the base level. Plastic activity thus tends to be
confined in the valley and can no longer fully explore the disordered landscape.

In Fig. \ref{ActivityMaps-vs-d0}, the values $\delta=0.5$ of the
age-like parameter has been used. From top to left, the evolution of
the plastic strain field is shown for values of the slip increment
$d_0=0.03,\;0.1,\;0.3$. A similar behavior as above is obtained. We
see that the higher value of the temperature like parameter $d_0$, the
less intense the localization and the faster the subsequent widening
process. Age- and mechanical temperature-like parameters $\delta$ and
$d_0$ thus seem to behave as could be expected, at least
phenomenologically.

\section{Slow relaxation of shear-banding}

The residual stress field, is the self-balanced stress field which
results from the local slip events taking place from the initial
(stress free) state. The latter has a zero volume average.  It allows
one (in conjunction with the local random yield threshold) to
characterize the propensity of a site to undergo a plastic slip. This
motivates the recourse to standard tools used for aging behavior
characterization.  Two-point correlation functions based on the
residual stress field are proposed: 
\begin{widetext}
    \be
    C_{\sigma}(\varepsilon_w, \varepsilon_p)= \frac{\langle
    \sigma_{res}(\varepsilon_w,x)\sigma_{res}(\varepsilon_p,x)\rangle_x}
    {\left(\langle
    \sigma_{res}(\varepsilon_w,x)\sigma_{res}(\varepsilon_w,x)\rangle_x
    \langle
    \sigma_{res}(\varepsilon_p,x)\sigma_{res}(\varepsilon_p,x)\rangle_x
    \right)^{1/2}}
    \ee
\end{widetext}
where the symbol $\langle ...\rangle_x$ designates a spatial average over $x$. Note that the model does not
depend on time as such; the global plastic strain plays the role of an
evolution parameter.

\begin{figure*}[tb]
   \centering
\begin{minipage}[h]{1.0\textwidth}
\includegraphics[width=0.32\textwidth]{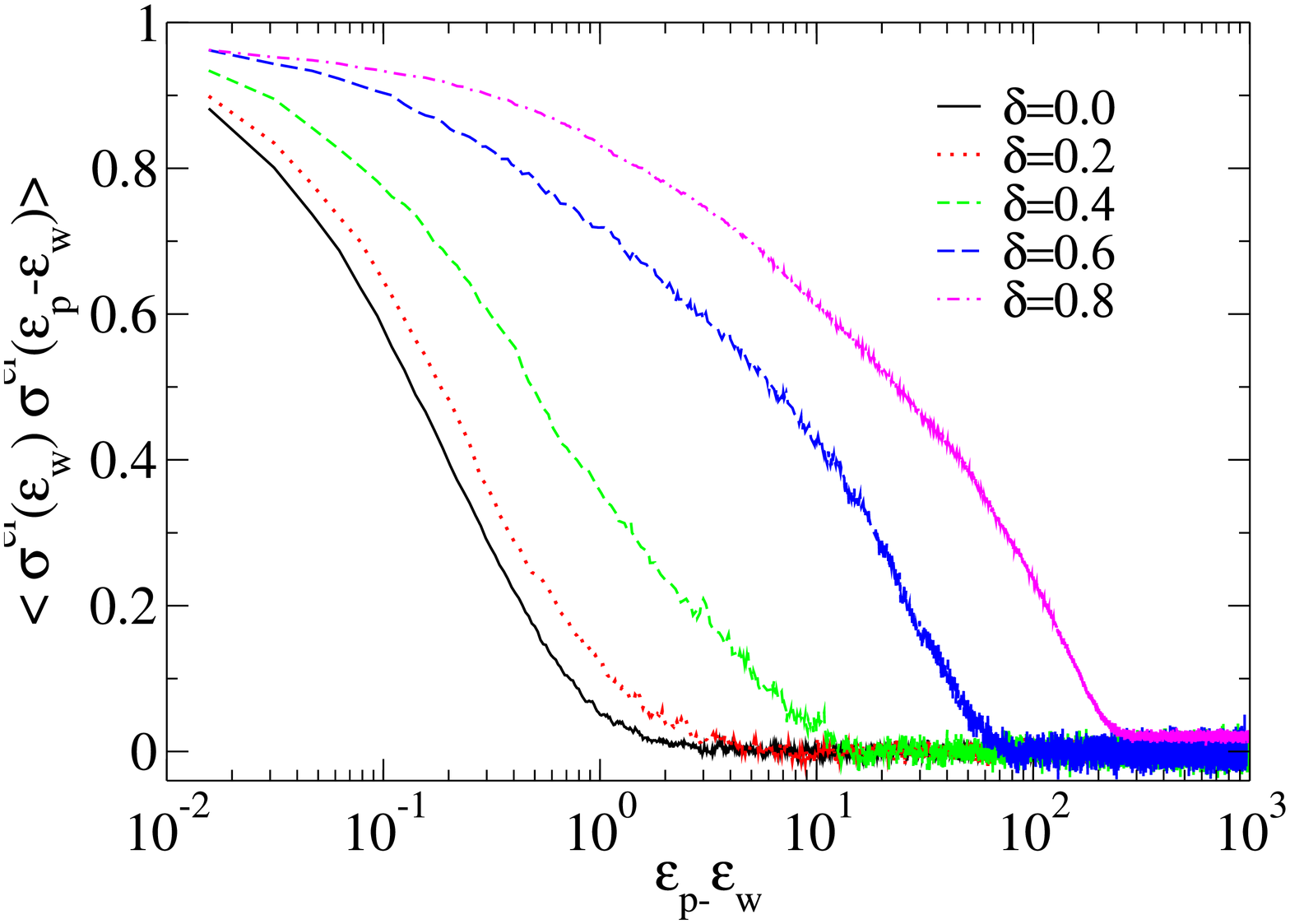}
\includegraphics[width=0.32\textwidth]{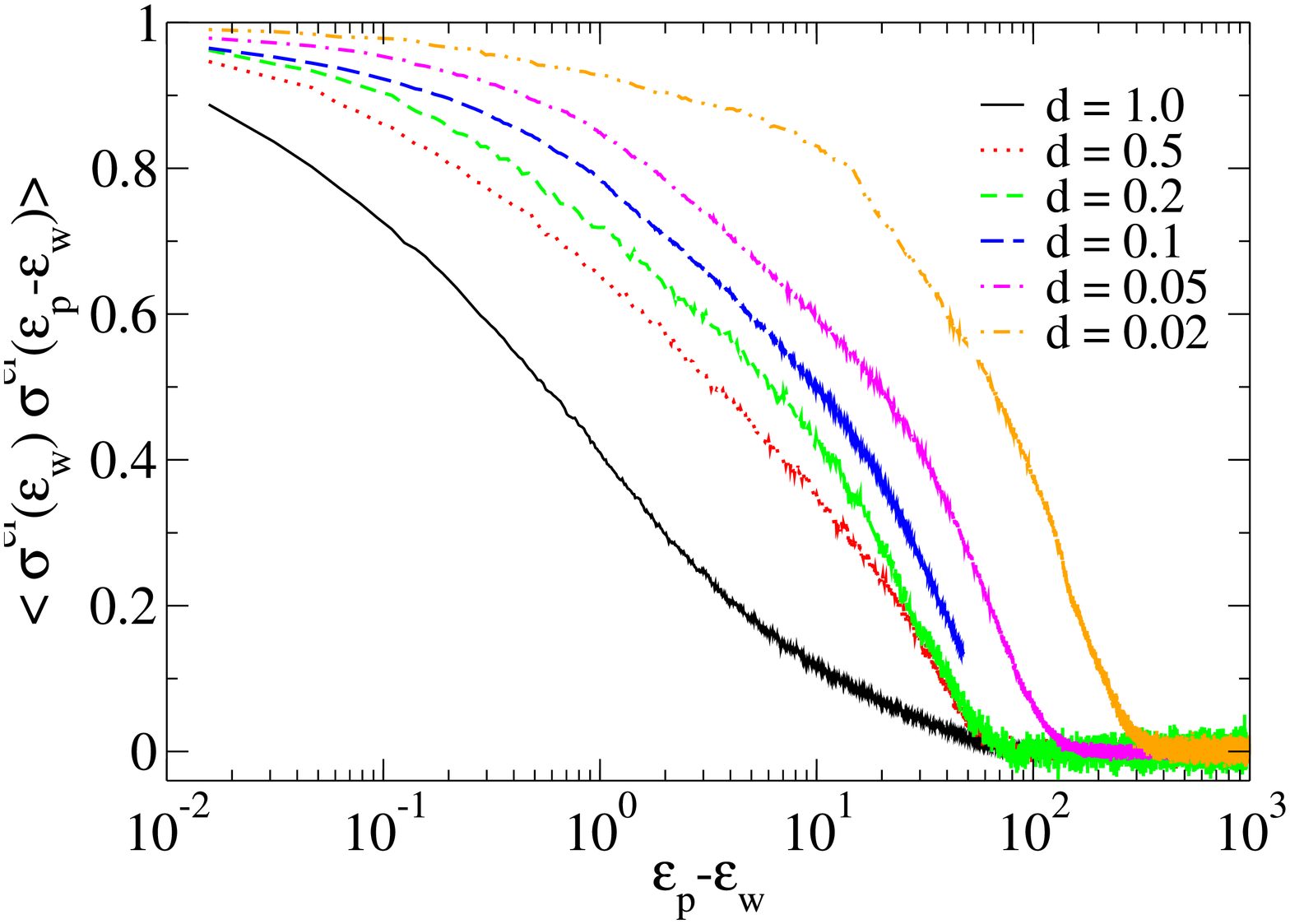}
\includegraphics[width=0.32\textwidth]{tau-vs-d-delta.eps}
\end{minipage}
\caption{Effect of the age-like parameter $\delta$ (Left) and of the
mechanical noise amplitude $d_0$ (Center) on the two-point stress
correlation function with $\varepsilon_w=1$.  (Right) Dependence on
$\delta$ and  $d_0$ of the typical plastic deformation $\varepsilon^*$
needed to relax shear banding. After rescaling data can be reasonably
collapsed onto a single master curve; the dashed line indicates an
exponential behavior accounting for the glassiness of shear-banding
at high $\delta$ \--- low $d_0$. }
  \label{cor-vs-d0-delta}
\end{figure*}

Such two-point correlation functions can be used to follow the formation and the
subsequent relaxation of shear-banding\cite{VR-preprint11}. In the following we
only discuss the relaxation stage after the initial transient and full formation
of the shear band. In Fig. \ref{cor-vs-d0-delta} we present the dependence of
the stress correlation for $\varepsilon_w=1$; at this deformation level, which
corresponds to the typical amplitude of the local plastic threshold,
localization (if any) is fully set.

The left panel of Fig. \ref{cor-vs-d0-delta} shows the effect of
the ``age'' parameter, with $\delta=0,\;0.2,\;0.4\;,0.6\;,0.8$ and
$d_0=0.2$. In the un-aged configuration ($\delta=0$), the system
decorrelates after a typical plastic strain
$\varepsilon_p=d_0^{0.5}$. This reflects the
non persistence of localization in the standard un-aged case. In the
case of an aged initial configuration we obtain significantly
different results. The systems appears to decorrelate only after a
plastic deformation growing exponentially with the parameter
$\delta$. Moreover when fitting data with a simple stretched
exponential, the exponent can be shown to transit from values
slightly below unity in the un-aged case to values close or below
$1/2$ in the more aged configurations. The shear banding persistence
thus seems to directly depend on the age.

Pursuing the above discussed analogy we now show in the center panel
of Fig. \ref{cor-vs-d0-delta} the correlation functions obtained
with a fixed age parameter $\delta=0.6$ for values of the slip
increment parameter varying from $d_0=0.02$ to $d_0=1$ (computations
were performed on lattices of size $64\times64$ with 20 to 200
realizations). As could be anticipated from the above displayed maps
of plastic deformation, the shear-banding persistence tends to
increase inversely with the slip increment parameter, the lower $d_0$
the higher the decorrelation time. The slip increment parameter $d_0$
thus seems reasonably to act as the amplitude of a mechanical noise
allowing the system to escape its trapped state. In other words, $d_0$
which stands here for the product of the volume of a flipping zone
times its typical plastic strain seems to be a good candidate for the
elusive effective mechanical temperature discussed in the SGR
model\cite{Sollich-PRL97}.

We try to rationalize in the right panel of
Fig. \ref{cor-vs-d0-delta} the age and mechanical noise dependence
of the shear-banding persistence. Exploring the two-dimensional
space of parameters $\delta$ and $d_0$, using a simple stretched
exponential fitting procedure, we extracted the typical plastic
strain $\varepsilon^*$ associated with stress decorrelation after
shear-banding formation ($\varepsilon_w=1$ in the above
notations). This allows us to propose a reasonable scaling
dependence:
\be \varepsilon^* = d_0^a \varphi \left( \frac{\delta}{d_0^b}\right)\;,
\ee
where
$$
 \quad \varphi(x\to 0) \approx A
\;{\rm and} \quad \varphi(x\to \infty) \approx C.e^{Bx} \;.
$$
The choice $a=0.5$ and $b=0.2$ allowed us to obtain a reasonable collapse
of the data collected for $d_0 \in [0.02,0.03,0.05,0.1,0.2,0.5,1]$ and $\delta
\in [0,0.1,0.2,0.3,0.4,0.5,0.6,0.7,0.8,0.9]$. In Fig. \ref{cor-vs-d0-delta} an
indicative exponential curve is shown to account for the high age and/or low
mechanical noise shear banding slow relaxation behavior.

\section{Conclusion}

To summarize, we showed that our simple Eshelby-like mesoscopic model
of amorphous plasticity exhibits a striking dependence on initial
conditions. The introduction of a simple bias to shift the initial
distribution of local yield stress values from its counterpart used to
renew the yield stress after local reorganization has a remarkable
consequence: the systems self-traps in a localized state to form a
shear band and remains so for a longer and longer ``time'' when the
bias value increases. This bias can thus be interpreted as an
estimator of the age of the system before shearing (or related to some
effective structural temperature\cite{Bouchbinder-PRE09b}.

Moreover we show that the ratio of the typical slip increment (more
rigorously in the formalism of the Eshelby inclusion, the volume of a
reorganizing zone times its typical plastic strain) on the typical
plastic yield stress acts as an effective mechanical temperature in
the sense proposed in the SGR model of Sollich et
al.\cite{Sollich-PRL97}. This parameter indeed gives the amplitude of
the mechanical noise induced by successive reorganizations. The lower
this amplitude, the longer the systems gets trapped and the slower the
widening of the shear bands.

In conclusion the present depinning model of amorphous
  plasticity appears to reproduce shear-banding, a crucial feature of
  the phenomenology of metallic glasses. The nucleation step of the
  shear-band is followed by a slow broadening step of the band. The
  latter is quantitatively characterized by a slow relaxation of the
  stress-stress correlation. Note that this slow dynamics
  spontaneously emerges in absence of any prescribed internal
  relaxation timescale.

The behavior of the model is controlled by two parameters that can be
associated to a {\sl structural} effective
temperature\cite{Bouchbinder-PRE09b}) and a {\sl mechanical} effective
temperature\cite{Sollich-PRL97} respectively. This model may thus
contribute to clarify the respective effects of structural relaxation
and mechanical noise induced by local reorganizations in the plastic
behavior of amorphous materials.

\begin{acknowledgments}
D.V. would like to thank M.L. Falk for enlightening discussions.
\end{acknowledgments}


\begin{thebibliography}{0}%
\makeatletter
\providecommand \@ifxundefined [1]{%
 \@ifx{#1\undefined}
}%
\providecommand \@ifnum [1]{%
 \ifnum #1\expandafter \@firstoftwo
 \else \expandafter \@secondoftwo
 \fi
}%
\providecommand \@ifx [1]{%
 \ifx #1\expandafter \@firstoftwo
 \else \expandafter \@secondoftwo
 \fi
}%
\providecommand \natexlab [1]{#1}%
\providecommand \enquote  [1]{``#1''}%
\providecommand \bibnamefont  [1]{#1}%
\providecommand \bibfnamefont [1]{#1}%
\providecommand \citenamefont [1]{#1}%
\providecommand \href@noop [0]{\@secondoftwo}%
\providecommand \href [0]{\begingroup \@sanitize@url \@href}%
\providecommand \@href[1]{\@@startlink{#1}\@@href}%
\providecommand \@@href[1]{\endgroup#1\@@endlink}%
\providecommand \@sanitize@url [0]{\catcode `\\12\catcode `\$12\catcode
  `\&12\catcode `\#12\catcode `\^12\catcode `\_12\catcode `\%12\relax}%
\providecommand \@@startlink[1]{}%
\providecommand \@@endlink[0]{}%
\providecommand \url  [0]{\begingroup\@sanitize@url \@url }%
\providecommand \@url [1]{\endgroup\@href {#1}{\urlprefix }}%
\providecommand \urlprefix  [0]{URL }%
\providecommand \Eprint [0]{\href }%
\@ifxundefined \urlstyle {%
  \providecommand \doi  [0]{\begingroup \@sanitize@url \@doi}%
  \providecommand \@doi [1]{\endgroup \@@startlink {\doibase
  #1}doi:\discretionary {}{}{}#1\@@endlink }%
}{%
  \providecommand \doi  [0]{doi:\discretionary{}{}{}\begingroup
  \urlstyle{rm}\Url }%
}%
\providecommand \doibase [0]{http://dx.doi.org/}%
\providecommand \Doi [0]{\begingroup \@sanitize@url \@Doi }%
\providecommand \@Doi  [1]{\endgroup\@@startlink{\doibase#1}\@@Doi}%
\providecommand \@@Doi [1]{#1\@@endlink}%
\providecommand \selectlanguage [0]{\@gobble}%
\providecommand \bibinfo  [0]{\@secondoftwo}%
\providecommand \bibfield  [0]{\@secondoftwo}%
\providecommand \translation [1]{[#1]}%
\providecommand \BibitemOpen [0]{}%
\providecommand \bibitemStop [0]{}%
\providecommand \bibitemNoStop [0]{.\EOS\space}%
\providecommand \EOS [0]{\spacefactor3000\relax}%
\providecommand \BibitemShut  [1]{\csname bibitem#1\endcsname}%
\end{thebibliography}%


\begin{thebibliography}{10}%
\makeatletter
\providecommand \@ifxundefined [1]{%
 \ifx #1\undefined \expandafter \@firstoftwo
 \else \expandafter \@secondoftwo
\fi
}%
\providecommand \@ifnum [1]{%
 \ifnum #1\expandafter \@firstoftwo
 \else \expandafter \@secondoftwo
\fi
}%
\providecommand \enquote [1]{``#1''}%
\providecommand \bibnamefont  [1]{#1}%
\providecommand \bibfnamefont [1]{#1}%
\providecommand \citenamefont [1]{#1}%
\providecommand\href[0]{\@sanitize\@href}%
\providecommand\@href[1]{\endgroup\@@startlink{#1}\endgroup\@@href}%
\providecommand\@@href[1]{#1\@@endlink}%
\providecommand \@sanitize [0]{\begingroup\catcode`\&12\catcode`\#12\relax}%
\@ifxundefined \pdfoutput {\@firstoftwo}{%
 \@ifnum{\z@=\pdfoutput}{\@firstoftwo}{\@secondoftwo}%
}{%
 \providecommand\@@startlink[1]{\leavevmode}%
 \providecommand\@@endlink[0]{}%
}{%
 \providecommand\@@startlink[1]{%
  \leavevmode
  \pdfstartlink
   attr{/Border[0 0 1 ]/H/I/C[0 1 1]}%
   user{/Subtype/Link/A<</Type/Action/S/URI/URI(#1)>>}%
  \relax
 }%
 \providecommand\@@endlink[0]{\pdfendlink}%
}%
\providecommand \url  [0]{\begingroup\@sanitize \@url }%
\providecommand \@url [1]{\endgroup\@href {#1}{\urlprefix}}%
\providecommand \urlprefix [0]{URL }%
\providecommand \Eprint[0]{\href }%
\@ifxundefined \urlstyle {%
  \providecommand \doi [1]{doi:\discretionary{}{}{}#1}%
}{%
  \providecommand \doi [0]{doi:\discretionary{}{}{}\begingroup
  \urlstyle{rm}\Url }%
}%
\providecommand \doibase [0]{http://dx.doi.org/}%
\providecommand \Doi[1]{\href{\doibase#1}}%
\providecommand \bibAnnote [3]{%
  \BibitemShut{#1}%
  \begin{quotation}\noindent
    \textsc{Key:}\ #2\\\textsc{Annotation:}\ #3%
  \end{quotation}%
}%
\providecommand \bibAnnoteFile [2]{%
  \IfFileExists{#2}{\bibAnnote {#1} {#2} {\input{#2}}}{}%
}%
\providecommand \typeout [0]{\immediate \write \m@ne }%
\providecommand \selectlanguage [0]{\@gobble}%
\providecommand \bibinfo [0]{\@secondoftwo}%
\providecommand \bibfield [0]{\@secondoftwo}%
\providecommand \translation [1]{[#1]}%
\providecommand \BibitemOpen[0]{}%
\providecommand \bibitemStop [0]{}%
\providecommand \bibitemNoStop [0]{.\EOS\space}%
\providecommand \EOS [0]{\spacefactor3000\relax}%
\providecommand \BibitemShut [1]{\csname bibitem#1\endcsname}%
\bibitem{Schuh-ActaMat07}%
  \BibitemOpen
  \bibfield{author}{%
  \bibinfo {author} {\bibfnamefont{C.~A.}\ \bibnamefont{Schuh}}, \bibinfo
  {author} {\bibfnamefont{T.~C.}\ \bibnamefont{Hufnagel}},\ and\ \bibinfo
  {author} {\bibfnamefont{U.}~\bibnamefont{Ramamurty}},\ }%
  \bibfield{journal}{%
  \bibinfo {journal} {Acta. Mat.}\ }%
  \textbf{\bibinfo {volume} {55}},\ \bibinfo {pages} {4067} (\bibinfo {year}
  {2007})%
  \bibAnnoteFile{NoStop}{Schuh-ActaMat07}%
\bibitem{RTV-MSMSE11}%
  \BibitemOpen
  \bibfield{author}{%
  \bibinfo {author} {\bibfnamefont{D.}~\bibnamefont{Rodney}}, \bibinfo {author}
  {\bibfnamefont{A.}~\bibnamefont{Tanguy}},\ and\ \bibinfo {author}
  {\bibfnamefont{D.}~\bibnamefont{Vandembroucq}},\ }%
  \bibfield{journal}{%
  \bibinfo {journal} {submitted to Modell. Simul. Mater. Sci. Eng.},\ \bibinfo
  {pages} {arXiv:1107.2022}}%
   (\bibinfo {year} {2011})%
  \bibAnnoteFile{NoStop}{RTV-MSMSE11}%
\bibitem{Lewandowski-NatMat06}%
  \BibitemOpen
  \bibfield{author}{%
  \bibinfo {author} {\bibfnamefont{J.~J.}\ \bibnamefont{Lewandowski}}\ and\
  \bibinfo {author} {\bibfnamefont{A.~L.}\ \bibnamefont{Greer}},\ }%
  \bibfield{journal}{%
  \bibinfo {journal} {Nat. Mater.}\ }%
  \textbf{\bibinfo {volume} {5}},\ \bibinfo {pages} {15} (\bibinfo {year}
  {2006})%
  \bibAnnoteFile{NoStop}{Lewandowski-NatMat06}%
\bibitem{Falk-PRL05}%
  \BibitemOpen
  \bibfield{author}{%
  \bibinfo {author} {\bibfnamefont{Y.~F.}\ \bibnamefont{Shi}}\ and\ \bibinfo
  {author} {\bibfnamefont{M.~L.}\ \bibnamefont{Falk}},\ }%
  \bibfield{journal}{%
  \bibinfo {journal} {Phys. Rev. Lett.}\ }%
  \textbf{\bibinfo {volume} {95}},\ \bibinfo {pages} {095502} (\bibinfo {year}
  {2005})%
  \bibAnnoteFile{NoStop}{Falk-PRL05}%
\bibitem{Falk-PRB06}%
  \BibitemOpen
  \bibfield{author}{%
  \bibinfo {author} {\bibfnamefont{Y.~F.}\ \bibnamefont{Shi}}\ and\ \bibinfo
  {author} {\bibfnamefont{M.~L.}\ \bibnamefont{Falk}},\ }%
  \bibfield{journal}{%
  \bibinfo {journal} {Phys. Rev. B}\ }%
  \textbf{\bibinfo {volume} {73}},\ \bibinfo {pages} {214201} (\bibinfo {year}
  {2006})%
  \bibAnnoteFile{NoStop}{Falk-PRB06}%
\bibitem{Falk-PRL07}%
  \BibitemOpen
  \bibfield{author}{%
  \bibinfo {author} {\bibfnamefont{Y.~F.}\ \bibnamefont{Shi}}, \bibinfo
  {author} {\bibfnamefont{M.~B.}\ \bibnamefont{Katz}}, \bibinfo {author}
  {\bibfnamefont{H.}~\bibnamefont{Li}},\ and\ \bibinfo {author}
  {\bibfnamefont{M.~L.}\ \bibnamefont{Falk}},\ }%
  \bibfield{journal}{%
  \bibinfo {journal} {Phys. Rev. Lett.}\ }%
  \textbf{\bibinfo {volume} {98}},\ \bibinfo {pages} {185505} (\bibinfo {year}
  {2007})%
  \bibAnnoteFile{NoStop}{Falk-PRL07}%
\bibitem{FalkLanger-PRE98}%
  \BibitemOpen
  \bibfield{author}{%
  \bibinfo {author} {\bibfnamefont{M.~L.}\ \bibnamefont{Falk}}\ and\ \bibinfo
  {author} {\bibfnamefont{J.~S.}\ \bibnamefont{Langer}},\ }%
  \bibfield{journal}{%
  \bibinfo {journal} {Phys. Rev. E}\ }%
  \textbf{\bibinfo {volume} {57}},\ \bibinfo {pages} {7192} (\bibinfo {year}
  {1998})%
  \bibAnnoteFile{NoStop}{FalkLanger-PRE98}%
\bibitem{Argon-ActaMet79}%
  \BibitemOpen
  \bibfield{author}{%
  \bibinfo {author} {\bibfnamefont{A.~S.}\ \bibnamefont{Argon}},\ }%
  \bibfield{journal}{%
  \bibinfo {journal} {Acta Metall.}\ }%
  \textbf{\bibinfo {volume} {27}},\ \bibinfo {pages} {47} (\bibinfo {year}
  {1979})%
  \bibAnnoteFile{NoStop}{Argon-ActaMet79}%
\bibitem{Manning-PRE07}%
  \BibitemOpen
  \bibfield{author}{%
  \bibinfo {author} {\bibfnamefont{M.~L.}\ \bibnamefont{Manning}}, \bibinfo
  {author} {\bibfnamefont{J.~S.}\ \bibnamefont{Langer}},\ and\ \bibinfo
  {author} {\bibfnamefont{J.~M.}\ \bibnamefont{Carlson}},\ }%
  \bibfield{journal}{%
  \bibinfo {journal} {Phys. Rev. E}\ }%
  \textbf{\bibinfo {volume} {76}},\ \bibinfo {pages} {056106} (\bibinfo {year}
  {2007})%
  \bibAnnoteFile{NoStop}{Manning-PRE07}%
\bibitem{Manning-PRE09}%
  \BibitemOpen
  \bibfield{author}{%
  \bibinfo {author} {\bibfnamefont{M.~L.}\ \bibnamefont{Manning}}, \bibinfo
  {author} {\bibfnamefont{E.~G.}\ \bibnamefont{Daub}}, \bibinfo {author}
  {\bibfnamefont{J.~S.}\ \bibnamefont{Langer}},\ and\ \bibinfo {author}
  {\bibfnamefont{J.~M.}\ \bibnamefont{Carlson}},\ }%
  \bibfield{journal}{%
  \bibinfo {journal} {Phys. Rev. E}\ }%
  \textbf{\bibinfo {volume} {79}},\ \bibinfo {pages} {016110} (\bibinfo {year}
  {2009})%
  \bibAnnoteFile{NoStop}{Manning-PRE09}%
\bibitem{Bouchaud-JPI92}%
  \BibitemOpen
  \bibfield{author}{%
  \bibinfo {author} {\bibfnamefont{J.~P.}\ \bibnamefont{Bouchaud}},\ }%
  \bibfield{journal}{%
  \bibinfo {journal} {J. Phys. I}\ }%
  \textbf{\bibinfo {volume} {2}},\ \bibinfo {pages} {1705} (\bibinfo {year}
  {1992})%
  \bibAnnoteFile{NoStop}{Bouchaud-JPI92}%
\bibitem{Sollich-PRL97}%
  \BibitemOpen
  \bibfield{author}{%
  \bibinfo {author} {\bibfnamefont{P.}~\bibnamefont{Sollich}}, \bibinfo
  {author} {\bibfnamefont{F.}~\bibnamefont{Lequeux}}, \bibinfo {author}
  {\bibfnamefont{P.}~\bibnamefont{H\'ebraud}},\ and\ \bibinfo {author}
  {\bibfnamefont{M.~E.}\ \bibnamefont{Cates}},\ }%
  \bibfield{journal}{%
  \bibinfo {journal} {Phys. Rev. Lett.}\ }%
  \textbf{\bibinfo {volume} {78}},\ \bibinfo {pages} {2020} (\bibinfo {year}
  {1997})%
  \bibAnnoteFile{NoStop}{Sollich-PRL97}%
\bibitem{Eshelby57}%
  \BibitemOpen
  \bibfield{author}{%
  \bibinfo {author} {\bibfnamefont{J.~D.}\ \bibnamefont{Eshelby}},\ }%
  \bibfield{journal}{%
  \bibinfo {journal} {Proc. Roy. Soc. A}\ }%
  \textbf{\bibinfo {volume} {241}},\ \bibinfo {pages} {376} (\bibinfo {year}
  {1957})%
  \bibAnnoteFile{NoStop}{Eshelby57}%
\bibitem{BulatovArgon94a}%
  \BibitemOpen
  \bibfield{author}{%
  \bibinfo {author} {\bibfnamefont{V.~V.}\ \bibnamefont{Bulatov}}\ and\
  \bibinfo {author} {\bibfnamefont{A.~S.}\ \bibnamefont{Argon}},\ }%
  \bibfield{journal}{%
  \bibinfo {journal} {Modell. Simul. Mater. Sci. Eng.}\ }%
  \textbf{\bibinfo {volume} {2}},\ \bibinfo {pages} {167} (\bibinfo {year}
  {1994})%
  \bibAnnoteFile{NoStop}{BulatovArgon94a}%
\bibitem{BulatovArgon94b}%
  \BibitemOpen
  \bibfield{author}{%
  \bibinfo {author} {\bibfnamefont{V.~V.}\ \bibnamefont{Bulatov}}\ and\
  \bibinfo {author} {\bibfnamefont{A.~S.}\ \bibnamefont{Argon}},\ }%
  \bibfield{journal}{%
  \bibinfo {journal} {Modell. Simul. Mater. Sci. Eng.}\ }%
  \textbf{\bibinfo {volume} {2}},\ \bibinfo {pages} {185} (\bibinfo {year}
  {1994})%
  \bibAnnoteFile{NoStop}{BulatovArgon94b}%
\bibitem{BulatovArgon94c}%
  \BibitemOpen
  \bibfield{author}{%
  \bibinfo {author} {\bibfnamefont{V.~V.}\ \bibnamefont{Bulatov}}\ and\
  \bibinfo {author} {\bibfnamefont{A.~S.}\ \bibnamefont{Argon}},\ }%
  \bibfield{journal}{%
  \bibinfo {journal} {Modell. Simul. Mater. Sci. Eng.}\ }%
  \textbf{\bibinfo {volume} {2}},\ \bibinfo {pages} {203} (\bibinfo {year}
  {1994})%
  \bibAnnoteFile{NoStop}{BulatovArgon94c}%
\bibitem{BVR-PRL02}%
  \BibitemOpen
  \bibfield{author}{%
  \bibinfo {author} {\bibfnamefont{J.-C.}\ \bibnamefont{Baret}}, \bibinfo
  {author} {\bibfnamefont{D.}~\bibnamefont{Vandembroucq}},\ and\ \bibinfo
  {author} {\bibfnamefont{S.}~\bibnamefont{Roux}},\ }%
  \bibfield{journal}{%
  \bibinfo {journal} {Phys. Rev. Lett.}\ }%
  \textbf{\bibinfo {volume} {89}},\ \bibinfo {pages} {195506} (\bibinfo {year}
  {2002})%
  \bibAnnoteFile{NoStop}{BVR-PRL02}%
\bibitem{Picard-PRE02}%
  \BibitemOpen
  \bibfield{author}{%
  \bibinfo {author} {\bibfnamefont{G.}~\bibnamefont{Picard}}, \bibinfo {author}
  {\bibfnamefont{A.}~\bibnamefont{Ajdari}}, \bibinfo {author}
  {\bibfnamefont{L.}~\bibnamefont{Bocquet}},\ and\ \bibinfo {author}
  {\bibfnamefont{F.}~\bibnamefont{Lequeux}},\ }%
  \bibfield{journal}{%
  \bibinfo {journal} {Phys. Rev. E}\ }%
  \textbf{\bibinfo {volume} {66}},\ \bibinfo {pages} {051501} (\bibinfo {year}
  {2002})%
  \bibAnnoteFile{NoStop}{Picard-PRE02}%
\bibitem{Picard-PRE05}%
  \BibitemOpen
  \bibfield{author}{%
  \bibinfo {author} {\bibfnamefont{G.}~\bibnamefont{Picard}}, \bibinfo {author}
  {\bibfnamefont{A.}~\bibnamefont{Ajdari}}, \bibinfo {author}
  {\bibfnamefont{F.}~\bibnamefont{Lequeux}},\ and\ \bibinfo {author}
  {\bibfnamefont{L.}~\bibnamefont{Bocquet}},\ }%
  \bibfield{journal}{%
  \bibinfo {journal} {Phys. Rev. E}\ }%
  \textbf{\bibinfo {volume} {71}},\ \bibinfo {pages} {010501(R)} (\bibinfo
  {year} {2005})%
  \bibAnnoteFile{NoStop}{Picard-PRE05}%
\bibitem{Jagla-PRE07}%
  \BibitemOpen
  \bibfield{author}{%
  \bibinfo {author} {\bibfnamefont{E.~A.}\ \bibnamefont{Jagla}},\ }%
  \bibfield{journal}{%
  \bibinfo {journal} {Phys. Rev. E}\ }%
  \textbf{\bibinfo {volume} {76}},\ \bibinfo {pages} {046119} (\bibinfo {year}
  {2007})%
  \bibAnnoteFile{NoStop}{Jagla-PRE07}%
\bibitem{Schuh-ActaMat09}%
  \BibitemOpen
  \bibfield{author}{%
  \bibinfo {author} {\bibfnamefont{E.~R.}\ \bibnamefont{Homer}}\ and\ \bibinfo
  {author} {\bibfnamefont{C.~A.}\ \bibnamefont{Schuh}},\ }%
  \bibfield{journal}{%
  \bibinfo {journal} {Acta Mat.}\ }%
  \textbf{\bibinfo {volume} {57}},\ \bibinfo {pages} {2823} (\bibinfo {year}
  {2009})%
  \bibAnnoteFile{NoStop}{Schuh-ActaMat09}%
\bibitem{Homer-PRB10}%
  \BibitemOpen
  \bibfield{author}{%
  \bibinfo {author} {\bibfnamefont{E.~R.}\ \bibnamefont{Homer}}, \bibinfo
  {author} {\bibfnamefont{D.}~\bibnamefont{Rodney}},\ and\ \bibinfo {author}
  {\bibfnamefont{C.~A.}\ \bibnamefont{Schuh}},\ }%
  \bibfield{journal}{%
  \bibinfo {journal} {Phys. Rev. B}\ }%
  \textbf{\bibinfo {volume} {81}},\ \bibinfo {pages} {064204} (\bibinfo {year}
  {2010})%
  \bibAnnoteFile{NoStop}{Homer-PRB10}%
\bibitem{TPRV-meso10}%
  \BibitemOpen
  \bibfield{author}{%
  \bibinfo {author} {\bibfnamefont{M.}~\bibnamefont{Talamali}}, \bibinfo
  {author} {\bibfnamefont{V.}~\bibnamefont{Pet\"aj\"a}}, \bibinfo {author}
  {\bibfnamefont{D.}~\bibnamefont{Vandembroucq}},\ and\ \bibinfo {author}
  {\bibfnamefont{S.}~\bibnamefont{Roux}},\ }%
  \bibfield{journal}{%
  \bibinfo {journal} {submitted to C.R. M\'ecanique}\ }%
  \textbf{\bibinfo {volume} {arXiv:1005.2463}} (\bibinfo {year} {2010})%
  \bibAnnoteFile{NoStop}{TPRV-meso10}%
\bibitem{TPVR-PRE11}%
  \BibitemOpen
  \bibfield{author}{%
  \bibinfo {author} {\bibfnamefont{M.}~\bibnamefont{Talamali}}, \bibinfo
  {author} {\bibfnamefont{V.}~\bibnamefont{Pet\"aj\"a}}, \bibinfo {author}
  {\bibfnamefont{D.}~\bibnamefont{Vandembroucq}},\ and\ \bibinfo {author}
  {\bibfnamefont{S.}~\bibnamefont{Roux}},\ }%
  \bibfield{journal}{%
  \bibinfo {journal} {Phys. Rev. E}\ }%
  \textbf{\bibinfo {volume} {84}},\ \bibinfo {pages} {016115} (\bibinfo {year}
  {2011})%
  \bibAnnoteFile{NoStop}{TPVR-PRE11}%
\bibitem{Maloney-JPCM08}%
  \BibitemOpen
  \bibfield{author}{%
  \bibinfo {author} {\bibfnamefont{C.~E.}\ \bibnamefont{Maloney}}\ and\
  \bibinfo {author} {\bibfnamefont{M.~O.}\ \bibnamefont{Robbins}},\ }%
  \bibfield{journal}{%
  \bibinfo {journal} {J. Phys. Cond. Matt.}\ }%
  \textbf{\bibinfo {volume} {20}},\ \bibinfo {pages} {244128} (\bibinfo {year}
  {2008})%
  \bibAnnoteFile{NoStop}{Maloney-JPCM08}%
\bibitem{Maloney-PRL09}%
  \BibitemOpen
  \bibfield{author}{%
  \bibinfo {author} {\bibfnamefont{C.~E.}\ \bibnamefont{Maloney}}\ and\
  \bibinfo {author} {\bibfnamefont{M.~O.}\ \bibnamefont{Robbins}},\ }%
  \bibfield{journal}{%
  \bibinfo {journal} {Phys. Rev. Lett.}\ }%
  \textbf{\bibinfo {volume} {102}},\ \bibinfo {pages} {225502} (\bibinfo {year}
  {2009})%
  \bibAnnoteFile{NoStop}{Maloney-PRL09}%
\bibitem{Fielding-SM09}%
  \BibitemOpen
  \bibfield{author}{%
  \bibinfo {author} {\bibfnamefont{S.~M.}\ \bibnamefont{Fielding}}, \bibinfo
  {author} {\bibfnamefont{M.~E.}\ \bibnamefont{Cates}},\ and\ \bibinfo {author}
  {\bibfnamefont{P.}~\bibnamefont{Sollich}},\ }%
  \bibfield{journal}{%
  \bibinfo {journal} {Soft Matter}\ }%
  \textbf{\bibinfo {volume} {5}},\ \bibinfo {pages} {2378} (\bibinfo {year}
  {2009})%
  \bibAnnoteFile{NoStop}{Fielding-SM09}%
\bibitem{Fielding-PRL11}%
  \BibitemOpen
  \bibfield{author}{%
  \bibinfo {author} {\bibfnamefont{R.~L.}\ \bibnamefont{Moorcroft}}, \bibinfo
  {author} {\bibfnamefont{M.~E.}\ \bibnamefont{Cates}},\ and\ \bibinfo {author}
  {\bibfnamefont{S.~M.}\ \bibnamefont{Fielding}},\ }%
  \bibfield{journal}{%
  \bibinfo {journal} {Phys. Rev. Lett.}\ }%
  \textbf{\bibinfo {volume} {106}},\ \bibinfo {pages} {055502} (\bibinfo {year}
  {2011})%
  \bibAnnoteFile{NoStop}{Fielding-PRL11}%
\bibitem{Lemaitre-preprint06}%
  \BibitemOpen
  \bibfield{author}{%
  \bibinfo {author} {\bibfnamefont{A.}~\bibnamefont{Lema{\^i}tre}}\ and\
  \bibinfo {author} {\bibfnamefont{C.}~\bibnamefont{Caroli}},\ }%
  \bibfield{journal}{%
  \bibinfo {journal} {arXiv:cond-mat/0609689}}%
   (\bibinfo {year} {2006})%
  \bibAnnoteFile{NoStop}{Lemaitre-preprint06}%
\bibitem{TPRV-PRE08}%
  \BibitemOpen
  \bibfield{author}{%
  \bibinfo {author} {\bibfnamefont{M.}~\bibnamefont{Talamali}}, \bibinfo
  {author} {\bibfnamefont{V.}~\bibnamefont{Pet\"aj\"a}}, \bibinfo {author}
  {\bibfnamefont{D.}~\bibnamefont{Vandembroucq}},\ and\ \bibinfo {author}
  {\bibfnamefont{S.}~\bibnamefont{Roux}},\ }%
  \bibfield{journal}{%
  \bibinfo {journal} {Phys. Rev. E}\ }%
  \textbf{\bibinfo {volume} {78}},\ \bibinfo {pages} {016109} (\bibinfo {year}
  {2008})%
  \bibAnnoteFile{NoStop}{TPRV-PRE08}%
\bibitem{Baumberger-AdvPhys06}%
  \BibitemOpen
  \bibfield{author}{%
  \bibinfo {author} {\bibfnamefont{T.}~\bibnamefont{Baumberger}}\ and\ \bibinfo
  {author} {\bibfnamefont{C.}~\bibnamefont{Caroli}},\ }%
  \bibfield{journal}{%
  \bibinfo {journal} {Adv. Phys.},\ \bibinfo {pages} {279}}%
   (\bibinfo {year} {2006})%
  \bibAnnoteFile{NoStop}{Baumberger-AdvPhys06}%
\bibitem{Behringer-GM10}%
  \BibitemOpen
  \bibfield{author}{%
  \bibinfo {author} {\bibfnamefont{J.}~\bibnamefont{Zhang}}, \bibinfo {author}
  {\bibfnamefont{T.}~\bibnamefont{Majmudar}}, \bibinfo {author}
  {\bibfnamefont{A.}~\bibnamefont{Tordesillas}},\ and\ \bibinfo {author}
  {\bibfnamefont{R.~B.}\ \bibnamefont{ger}},\ }%
  \bibfield{journal}{%
  \bibinfo {journal} {Granular Matter}\ }%
  \textbf{\bibinfo {volume} {12}},\ \bibinfo {pages} {159} (\bibinfo {year}
  {2010})%
  \bibAnnoteFile{NoStop}{Behringer-GM10}%
\bibitem{Rottler-PRL05}%
  \BibitemOpen
  \bibfield{author}{%
  \bibinfo {author} {\bibfnamefont{J.}~\bibnamefont{Rottler}}\ and\ \bibinfo
  {author} {\bibfnamefont{M.~O.}\ \bibnamefont{Robbins}},\ }%
  \bibfield{journal}{%
  \bibinfo {journal} {Phys. Rev. Lett.}\ }%
  \textbf{\bibinfo {volume} {95}},\ \bibinfo {pages} {225504} (\bibinfo {year}
  {2005})%
  \bibAnnoteFile{NoStop}{Rottler-PRL05}%
\bibitem{Bouchbinder-PRE09b}%
  \BibitemOpen
  \bibfield{author}{%
  \bibinfo {author} {\bibfnamefont{E.}~\bibnamefont{Bouchbinder}}\ and\
  \bibinfo {author} {\bibfnamefont{J.~S.}\ \bibnamefont{Langer}},\ }%
  \bibfield{journal}{%
  \bibinfo {journal} {Phys. Rev. E}\ }%
  \textbf{\bibinfo {volume} {80}},\ \bibinfo {pages} {031132} (\bibinfo {year}
  {2009})%
  \bibAnnoteFile{NoStop}{Bouchbinder-PRE09b}%
\bibitem{Torok-PRL00}%
  \BibitemOpen
  \bibfield{author}{%
  \bibinfo {author} {\bibfnamefont{J.}~\bibnamefont{T\"or\"ok}}, \bibinfo
  {author} {\bibfnamefont{S.}~\bibnamefont{Krishnamurthy}}, \bibinfo {author}
  {\bibfnamefont{J.}~\bibnamefont{Kert\'esz}},\ and\ \bibinfo {author}
  {\bibfnamefont{S.}~\bibnamefont{Roux}},\ }%
  \bibfield{journal}{%
  \bibinfo {journal} {Phys. Rev. Lett.}\ }%
  \textbf{\bibinfo {volume} {84}},\ \bibinfo {pages} {3851} (\bibinfo {year}
  {2000})%
  \bibAnnoteFile{NoStop}{Torok-PRL00}%
\bibitem{VR-preprint11}%
  \BibitemOpen
  \bibfield{author}{%
  \bibinfo {author} {\bibfnamefont{D.}~\bibnamefont{Vandembroucq}}\ and\
  \bibinfo {author} {\bibfnamefont{S.}~\bibnamefont{Roux}},\ }%
  \bibfield{journal}{%
  \bibinfo {journal} {unpublished}}%
   (\bibinfo {year} {2011})%
  \bibAnnoteFile{NoStop}{VR-preprint11}%
\end{thebibliography}

%

\end{document}